\documentclass[aps,prx,twocolumn,floatfix,superscriptaddress,longbibliography,nofootinbib]{revtex4-2}
\usepackage{comment}
\usepackage[utf8]{inputenc}
\usepackage{natbib}
\usepackage{xcolor}
\usepackage{mathtools}
\usepackage{amsmath}
\usepackage[colorlinks, linkcolor=blue]{hyperref}
\usepackage{amssymb}
\usepackage{gensymb}
\usepackage{graphicx}
\makeatletter
\@ifundefined{switch@array}{}{%
  \def\switch@array{}%
}
\makeatother

\usepackage{tabularx} 
\usepackage{tabularx}
\usepackage{epstopdf}
\usepackage{latexsym}
\usepackage{color, colortbl}
\usepackage{braket}
\usepackage{psfrag}
\usepackage{bbm}
\usepackage{bm}
\usepackage{titlesec}
\usepackage{dsfont}
\usepackage{multirow}
\usepackage{enumitem}
\usepackage[normalem]{ulem}
\usepackage{pdfpages}
\usepackage{xr}
\makeatletter
\AtBeginDocument{\let\LS@rot\@undefined}
\makeatother
\usepackage{tikz}
\renewcommand{\vec}[1]{\boldsymbol{#1}}
\def \nn {\nonumber}

\def \x {{\vec x}}

\def \beq {\begin{eqnarray}}
\def \eeq {\end{eqnarray}}

\def \tn {\textnormal}

\newcommand{\rU}{{\rm U}}
\newcommand{\rd}{{\rm d}}
\newcommand{\Fnorm}[1]{{\left\Vert {#1}\right\Vert}}
\newcommand{\calH} {{\mathcal{H}}}
\newcommand{\calQ} {{\mathcal{Q}}}
\newcommand{\Tr}{{\rm Tr}}
\newcommand{\Ad}{{\mathrm{Ad}}}
\newcommand{\AdH}{{\mathrm{Ad}_{H_1}}}

\begin{document}
\title{Floquet-Thermalization via Instantons near Dynamical Freezing}
\author{Rohit Mukherjee}\thanks{These authors contributed equally; Present address: Institute for Theoretical Physics, University of Cologne, 50937 Cologne, Germany}
\affiliation{Department of Physics, Cornell University, Ithaca, New York 14853, USA}
\author{Haoyu Guo}\thanks{These authors contributed equally}
\author{Debanjan Chowdhury} \email{debanjanchowdhury@cornell.edu}

\affiliation{Department of Physics, Cornell University, Ithaca, New York 14853, USA}

\begin{abstract}
    Periodically driven Floquet quantum many-body systems have revealed new insights into the rich interplay of thermalization, and growth of entanglement. The phenomenology of {\it dynamical freezing}, whereby a translationally invariant many-body system exhibits emergent conservation laws and a slow growth of entanglement entropy at certain fixed ratios of a drive amplitude and frequency, presents a novel paradigm for retaining memory of an initial state up to late times. Previous studies of dynamical freezing have largely been restricted to a high-frequency Floquet-Magnus expansion, and numerical exact diagonalization. Both techniques are unable to capture the slow approach to thermalization, or lack thereof, in a systematic fashion. 
    By employing Floquet {\it flow-renormalization}, where the time-dependent part of the Hamiltonian is gradually decoupled from the effective Hamiltonian using a sequence of unitary transformations, we unveil the universal approach to dynamical freezing and beyond, at asymptotically late times. We analyze the {\it fixed-point} behavior associated with the flow-renormalization at and near freezing using both exact-diagonalization and tensor-network based methods, and contrast the results with conventional prethermal phenomenon. 
    For a generic non-integrable spin Hamiltonian with a periodic cosine wave drive, the flow approaches an unstable fixed point with an approximate emergent symmetry. 
    We observe that at freezing the thermalization timescales are delayed compared to away from freezing, and the flow trajectory undergoes a series of {\it instanton} events. Our numerical results are supported by analytical solutions to the flow equations.
\end{abstract}
\maketitle
\section{Introduction}

A cornerstone of statistical mechanics of closed and chaotic systems is the ergodic hypothesis, whereby in the late-time limit the entire many-body phase-space is explored subject to the macroscopic constraints. This hypothesis underpins the equal probability assumption associated with the microcanonical ensemble, and ensures that the late-time average of observables reflect the microcanonical predictions. The generalization to quantum statistical mechanics, based on the Eigenstate Thermalization Hypothesis (ETH), provides a framework for thermalization in closed quantum systems evolving under their own unitary many-body dynamics~\cite{Deutsch,srednicki,Tasaki,rigol,d2016quantum}. As per ETH,  the eigenstates of generic quantum Hamiltonians exhibit ``typical" statistical properties, aligning closely with predictions from the microcanonical ensemble. Extending these concepts to periodically driven (i.e. Floquet) non-integrable quantum many-body problems without energy conservation, the natural fate is thermalization to an ``infinite-temperature" (i.e. featureless) state~\cite{RevModPhys.83.863,PhysRevX.4.041048}. A number of recent advances have revealed exciting mechanisms to evade the process of Floquet thermalization. One such possibility arises in strongly disordered Floquet many-body localized (MBL) systems~\cite{fmbl2,fmbl3,PhysRevLett.116.250401,PhysRevB.94.224202,PhysRevLett.124.190601,ABANIN20161,PhysRevB.93.174202}, building on progress in the study of MBL \cite{MBLn1,MBLn2,MBLn3,MBLn4,RMPMBL}, whose stability in the thermodynamic limit~\cite{mbltherm2,mbltherm1,sels,floquettherm3} at asymptotically late times remains a topic of active research.
\begin{figure}[htb!]
    \centering
    \includegraphics[width=1.0\linewidth]{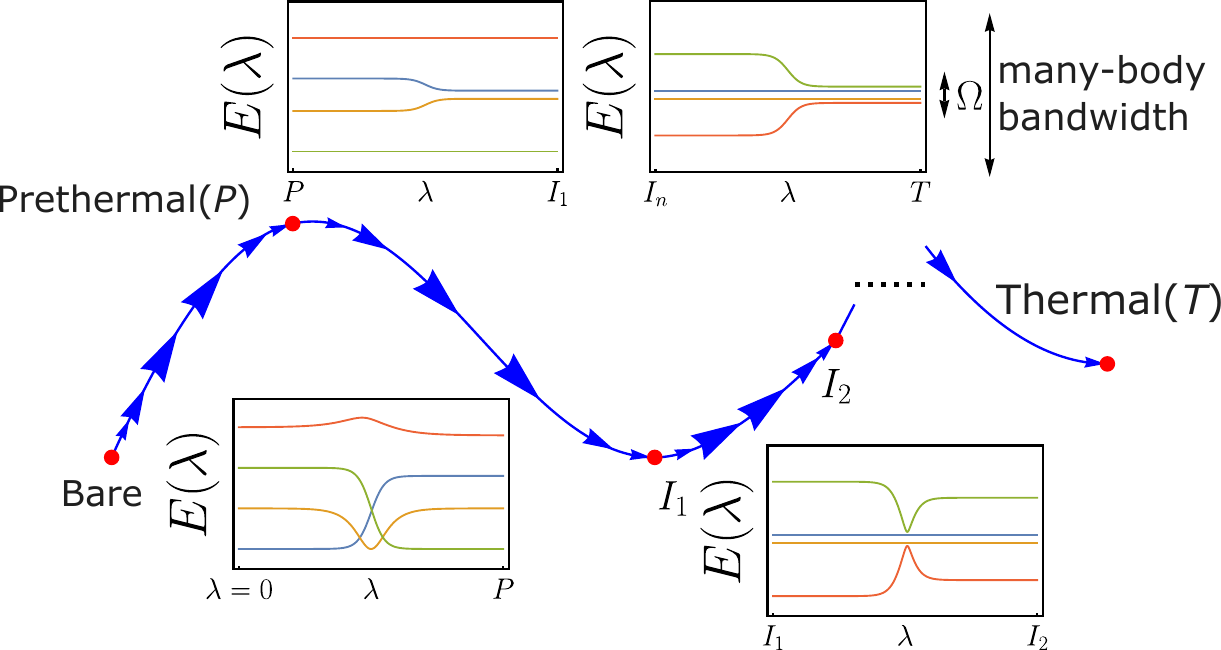}
    \caption{A schematic depiction of the Floquet flow-renormalization trajectory (blue) along the fRG time, $\lambda$. The insets show the evolution of many-body energy-levels between the different fixed points, e.g. from $\lambda=0$ to the first (unstable) prethermal fixed-point (`$P$'), followed by a sequence of instanton events that connect intermediate fixed points ($I_1,~I_2,...$), and final approach to a possibly thermal (`$T$') fixed-point, depending on a non-trivial interplay of the drive frequency and finite system size. The flow starts with the bare Hamiltonian (many-body bandwidth, $W$), and a Floquet drive with frequency, $\Omega$. The flow to $P$ leads to a renormalization of the many-body spectrum. The energy-levels associated with the effective Hamiltonian at the thermal fixed-point $T$ are folded into an interval smaller than $\Omega$. See Table~\ref{tab:timescales} for a relationship between the fRG time, $\lambda$, and real time, $t$, along with a summary of the different regimes encountered during the flow vis-\`a-vis these timescales.}
    \label{fig:Schematic}
\end{figure}
Another possibility are quantum scars~\cite{bluvstein2021controlling,serbyn2021quantum,Scar1}, which represent a unique but relatively rare class of excited states in the many-body spectrum that preserve memory of initial conditions. Finally, yet another mechanism invokes the phenomenon of Hilbert-space shattering~\cite{moudgalya2022quantum,PhysRevX.10.011047} for breaking ergodicity, which requires a set of conservation laws that fragment the Hilbert-space into exponentially many dynamically isolated subsectors, revealing dynamical behavior akin to both MBL and quantum scars~\cite{PhysRevB.101.174204}. At high drive frequencies, heating and thermalization can be significantly suppressed, leading to the emergence of prethermal plateaus~\cite{prethermal1,prthermal2,weidinger2017floquet,prethermal3,PhysRevLett.120.197601,prethermal4,prethermal6,prthermal5,PhysRevLett.132.100401}.\\

In parallel, a number of studies based primarily on numerical exact diagonalization (ED) and supporting Floquet-Magnus expansion, have reported the emergence of approximate conservation laws and slow dynamics in generic translationally invariant, periodically driven spin models at special ratios of the drive amplitude and frequency. This distinct route to avoiding Floquet thermalization has been denoted ``Dynamical Freezing"~\cite{ADas2010,PhysRevB.86.054410,PhysRevB.90.174407,PhysRevB.91.121106,PhysRevB.97.245122,KS2,AHaldar2021,KS7,freezingsree,KS1,KS4,KS5,tistadynamical,Roychowdhury_2024,AHaldar2024}. Recent works by some of us have extended this phenomenology to a strongly interacting fermionic setup analyzed using Keldysh field-theoretic methods~\cite{HGuo2024a}, and interacting bosonic models with a large local Hilbert-space dimension where a semiclassical correspondence reveals suppression of the chaotic Lyapunov exponent at the frozen points ~\cite{RMukherjee2024}. Unfortunately, a vast majority of analytical attempts to study dynamical freezing have relied on the {\it uncontrolled} and {\it perturbative} Floquet-Magnus expansion in a rotating frame, which does not produce a convergent series for generic interacting systems, and potentially misses important {\it non-perturbative} effects relevant for addressing fundamental questions related to thermalization (or lack thereof) at freezing.   This has left a number of basic questions about dynamical freezing for more generic drives unanswered, which will be addressed definitively in this work. Additionally, the numerical studies have largely been restricted to ED on relatively modest system-sizes upto finite times. The only notable exception is the recent numerical study for an infinite system size, but carried out for {\it only} a square-wave drive protocol \cite{AHaldar2024}. In contrast, the focus of this manuscript will be on the fate of a Floquet many-body system with a sinusoidal drive --- which is known to exhibit distinct behavior even within the perturbative Floquet-Magnus expansion \cite{RMukherjee2024} --- but now obtained using a number of complementary numerical and analytical non-pertubative methods.

First, it is natural to question what determines the location of the dynamically frozen points. For spin models with low-dimensional Hilbert spaces, the location typically correlates with the values of ratio of drive amplitude and frequency where the {\it leading-order} Magnus correction vanishes. However, some of us have previously demonstrated a quantitative shift in these locations due to the higher-order Magnus corrections in models with a larger on-site Hilbert-space dimension \cite{RMukherjee2024}. Finding an alternative method to determine the freezing points, without relying on a combination of numerics and the Floquet-Magnus expansion is a desirable goal. Second, and more importantly, whether dynamical freezing is {\it exact} and {\it robust} in the thermodynamic system-size and late-time limit, and if not, what controls the possible decay away from this special behavior has not been determined beyond the same limitations imposed by exact-diagonalization and Magnus-expansion for {\it generic} drives.

In this work, we improve on these limitations of prior studies and provide a universal description of dynamical freezing based on a Floquet flow-renormalization group (fRG) framework ~\cite{kehrein2007flow,MVogl2019,WG1,WG2,WG3,AP20,10.21468/SciPostPhys.11.2.028,MClaassen2021,thomson2024unravelling}. Fundamentally, the formalism generates a sequence of infinitesimal unitary transformations to systematically eliminate the time-dependent part of the driven Hamiltonian, resulting in a flow equation for the different coupling constants in the effective model. The method in the equilibrium setting is a generalization of the more widely known Schrieffer-Wolff transformations \cite{SW}; see Appendix \ref{Supp:SW} for a brief review of this connection and Appendix~\ref{app:numericaldetails} for details on its numerical implementation. As we will demonstrate in this paper by focusing on a generic, non-integrable one-dimensional spin model with a ``cosine-wave" drive, the flow trajectory along the fRG time at the {\it putative} freezing-point organizes into a sequence of {\it unstable} fixed-points, except the possibly inevitable thermal (`$T$') fixed-point that appears at asymptotically late times \footnote{For finite systems, this final fixed point appears ``thermal" only when the drive frequency $\Omega<\Omega_*(L)$, where $\Omega_*(L)$ is a system-size ($L$) dependent threshold frequency; see Sec.~\ref{sec:flow} and \ref{sec:thermalization} for more details. }; see Fig.~\ref{fig:Schematic} for a schematic depiction. Along the flow trajectory, the approach to the first fixed-point starting from the bare Hamiltonian is of special interest. It is an {\it unstable} prethermal (`$P$') fixed-point, associated with the emergence of an {\it approximately} conserved effective Hamiltonian, valid up to an exponentially long timescale in the drive frequency, $\Omega$. Subsequently, the system flows via a complex sequence of intermediate fixed points $I_1,I_2,\dots$. The flow in between these intermediate fixed points show characteristics of instanton events. However, as we will discuss below, it is important to note at the outset that capturing the flow to the thermal fixed-point numerically is exceptionally challenging given the exponential timescales involved, and remains an exciting arena for future research.

We will demonstrate explicitly that dynamical freezing manifests as an additional emergent conservation law associated with the prethermal fixed point. {\it Exact} dynamical freezing would correspond to a {\it vanishing} operator norm for the commutator between the emergent conserved quantity and the undriven part of the Hamiltonian near the prethermal fixed point. Generically, we find that the above commutator norm does {\it not} vanish exactly at freezing, but exhibits a non-trivial behavior as a function of the drive frequency, system size, and fRG time. Specifically, the emergent symmetry becomes more pronounced as the high frequency limit $\Omega\to\infty$ is approached. From that perspective, dynamical freezing is {\it asymptotic} in the high-frequency limit.

In this work, we have also analyzed carefully the flow awayfrom the unstable prethermal fixed point by combining powerful numerical and analytical techniques. Interestingly, we find that the system flows via a sequence of ``fixed-points", where the intermediate evolution in fRG time shows characteristics of instanton-like events. The renormalized effective Hamiltonian in the vicinity of these intermediate fixed-points is approximately constant and the associated flow is exponentially slow in $\exp(-\Omega/J)$, where $J$ is the typical single-particle energy scale of the system. The transition between two fixed points occurs rapidly in a ``short" window of fRG time, which is on the order of $J^{-1}$. We associate these ``tunneling" type events as instantons (see Fig.~\ref{fig:Schematic}), which are fundamentally non-perturbative and lie beyond the reach of more conventional perturbative methods, which includes the Floquet-Dyson perturbation theory \cite{AHaldar2021} or the Floquet-Magnus expansion~\cite{AEckardt2015}, but are readily accessible via the flow formalism. These perturbative methods are expected to describe the flow from the bare Hamiltonian to the prethermal fixed point. In contrast, the instantons signal a departure from the prethermal fixed point and the breakdown of perturbative expansions, which ultimately must capture the {\it universal} flow to the thermal fixed-point. We note that capturing and diagnosing the full flow towards this ``latest" time thermalization for any appreciable system size is an exceedingly slow process, given the exponentially long timescales involved, and is therefore well beyond the scope of any numerical method. However, as we discuss below, our results associated with the sequence of instanton-like events suggest that the ultimate fate and fixed-point behavior is tied to this slow thermalization for drive frequencies that lie below a system-size dependent threshold.

The rest of the paper is organized as follows: In Sec.~\ref{sec:flow}, we review the Floquet-flow formalism and the flow equation. We also introduce the characteristic features associated with dynamical freezing, as manifest in the flow formalism. In Sec.~\ref{sec:HO}, we use a simple example of a driven harmonic oscillator to illustrate the flow towards putative freezing using the flow formalism. In Sec.~\ref{sec:spinchain}, we turn to the solutions of the flow equations for a non-integrable spin chain driven by a cosine-wave first using both exact diagonalization and matrix-product operator based methods. In Sec.~\ref{sec:thermalization}, we present results for a number of numerical diagnostics that characterize the approach to thermalization near dynamical freezing as a function of system size and drive frequency. In Sec.~\ref{sec:analytics}, we turn to a general analytical treatment of the flow equation that applies to models beyond the specific spin chain Hamiltonian, which can describe the early-time approach to the unstable prethermal fixed-point, as well as the late-time thermalization due to instantons. We end with an outlook in Sec.~\ref{sec:outlook}. The appendices contain a number of technical details on the methods used in this paper. In App.~\ref{Supp:SW}, we review the connection between Schrieffer-Wolff transformations and the flow formalism. In App.~\ref{app:numericaldetails}, we describe some additional numerical details for simulating the flow equations. In App.~\ref{app:Supp}, we present some additional numerical results supporting conclusions in the main text. In App.~\ref{app:magnus}, we present details of analytical analysis of the flow equation in the early time regime, which behaves similarly as the Floquet-Magnus expansion. In App.~\ref{app:timescale}, we estimate the time scales associated with thermalization in the flow formalism. In App.~\ref{app:Magnus_spinchain}, we review the Floquet-Magnus expansion of the spin chain Hamiltonian we used.

 \section{Floquet-flow renormalization and \\dynamical Freezing}\label{sec:flow}
 We  review the flow formalism, developed originally by Wegner, Wilson and others \cite{wilsonwegner2,Glazek1993Phys.Rev.D} and adapted to a variety of strongly correlated problems \cite{kehrein2007flow} (including problems beyond the equilibrium setting \cite{WG1,WG2,WG3,MVogl2019,10.21468/SciPostPhys.11.2.028,MClaassen2021,thomson2024unravelling,10.21468/SciPostPhys.13.6.122,flowtime1,PhysRevB.106.115440}). The discussion here closely resembles the one introduced in Ref.~\cite{MClaassen2021}, which is reviewed in greater detail in Appendix \ref{Supp:SW}. 
 We are interested in time-dependent Hamiltonians expressed as,
\begin{equation}\label{eq:Hbare}
H(t)=H_0+H_1e^{i\Omega t}+H_{-1}e^{-i\Omega t},
\end{equation}
where $H_{-1}=H_1^\dagger$. Treating Eq.~\eqref{eq:Hbare} as the initial state of a renormalization process, the flow generates a sequence of infinitesimal unitary transformations, preserving the spectrum of the stroboscopic unitary evolution operator \cite{MBukov2015}, 
\begin{equation}\label{eq:exactU}
    U(T+t_0,t_0)=\mathcal{T} e^{-i\int_{t_0}^{T+t_0} \rd \tau H(\tau)}\,,
\end{equation}
where $\mathcal{T}$ denotes time-ordering. Our goal here is to realize a unitary transformation that systematically decouples the time-dependent driving component, $H_1$, from the effective theory, while at the same time preserve the spectrum of $U(T+t_0,t_0)$. Building on earlier work, the appropriate flow-equations are:
\begin{subequations}\label{eq:Flow}
\begin{eqnarray}
    \partial_\lambda H_0(\lambda)&=&2[H_1(\lambda),H_1^\dagger(\lambda)]\,, \label{eq:FlowH0} \\
     \partial_\lambda H_1(\lambda)&=&-\Omega H_1(\lambda)-\left[H_0(\lambda),H_1(\lambda)\right]\,,\label{eq:FlowH1}
\end{eqnarray}
\end{subequations}
where $\lambda$ represents the ``fRG time" along the flow trajectory. Note that $\lambda$ enters here dimensionally as real time  with dimension $\sim$[time] in contrast to the conventional Wegner-Wilson picture~\cite{wilsonwegner2,Glazek1993Phys.Rev.D}, where $\lambda \sim$ [time]$^2$. See Table~\ref{tab:timescales} for a glossary of the different timescales that describe the dynamics within the flow RG framework.

We solve the coupled matrix equations, Eq.~\eqref{eq:FlowH0} and Eq.~\eqref{eq:FlowH1}, using two approaches. First, for small system sizes we employ exact-diagonalization by representing the operators explicitly in the many-body Hilbert space and solving the equations exactly using the fourth-order Runge-Kutta (RK4) method~\cite{MVogl2019}. In the supplement \cite{supp}, we provide numerical evidences that Eq.~\eqref{eq:Flow} does preserve the eigenvalues of the Floquet Hamiltonian. The second approach is tailored for larger system sizes and utilizes matrix product operators (MPOs) \cite{MClaassen2021}. MPOs encode general many-body operators as products of operator-valued matrices, with each matrix acting locally at a single site. Once the operators $H_{0}(\lambda)$ and $H_{1}(\lambda)$ are expressed in the MPO form, we again solve the equations using the Runge-Kutta method. MPOs facilitate operations such as addition, evaluation, and computation of commutators, yielding exponentially localized MPOs~\cite{MPO1,MPO2}. MPOs have been applied to solve flow equations for many-body localized Hamiltonians~\cite{WG3,flowMBL1,flowMBL2}. All ED results are obtained using the QuSpin~\cite{PWeinberg2019,PWeinberg2017} Python package, while the MPO-based computations are performed using the Julia package ITensor~\cite{Julia-2017,fishman2022itensor,Itensor2}. See Appendix \ref{app:numericaldetails} for additional details.

We can view the non-oscillatory component, $H_0(\lambda)$, as an effective Hamiltonian responsible for generating the unitary evolution,
  \begin{equation}\label{eq:Ueff}
      U_\text{eff}(\lambda,t,t')=e^{-iH_0(\lambda)(t-t')}\,.
  \end{equation} The distance between $U_\text{eff}(\lambda,t_0+nT,t_0)$ and $U(t_0+nT,t_0)$ (Eq.~\eqref{eq:exactU}),  evaluated after $n$ cycles of the Floquet system and averaged over the starting time $t_0$ within one full cycle, is claimed to be bounded by the Frobenius norm of $nT\times H_{1}(\lambda)$~\cite{MClaassen2021}. 
  This implies that $U_\text{eff}$ serves as a reliable approximation for $U$ up to a time,
  \begin{equation}\label{eq:teff}
    t_\text{eff}(\lambda)\sim \left\Vert H_1(\lambda)\right\Vert^{-1} \Fnorm{\hat{1}},
  \end{equation} 
 which will provide useful estimates on the (pre-)thermalization timescales for our driven problem, as we will describe below. Here $\Fnorm{M}^2=\Tr(M^\dagger M)$ denotes the Frobenius norm of the operator, and Eq.~\eqref{eq:teff} is normalized by the norm of the identity operator $\hat{1}$ on the many-body Hilbert space.

  The typical behavior of the flow equation Eq.~\eqref{eq:Flow} can be divided into three regimes (see also Table~\ref{tab:timescales}): 
  \begin{itemize}
      \item[(a)] For fRG time, $0\leq\lambda\lesssim\Omega^{-1}$, the flow is in a ramp-up regime where $H_0(\lambda)$ is rapidly renormalized by the drive. In the high-frequency limit, this ramp-up flow can be described perturbatively using an expansion similar to the Floquet-Magnus expansion (see Sec.~\ref{sec:analytics}). 

      \item[(b)] Following this ramp-up regime, the system enters a prethermal regime where $\Omega^{-1}\lesssim \lambda \lesssim \lambda_\text{min}$. Here, the static part $H_0(\lambda)$ is approximately independent of $\lambda$, which leads to the interpretation that the system is close to the prethermal fixed point. The oscillatory part $H_1(\lambda)$ decays exponentially $\sim \exp(-\Omega\lambda)$ until $\lambda=\lambda_\text{min}$. Combining this with Eqs. \eqref{eq:Ueff} and \eqref{eq:teff} implies that in the prethermal regime the system shows approximate conservation of the Hamiltonian $H_0(\lambda_\text{min})$ up to an exponential long real time $t_\text{pre}\propto \exp(\Omega\lambda_\text{min})$. Here, the scale $\lambda_\text{min}$ is determined primarily by the local excitation energy scale (such as a single spin flip in a spin chain) of the system, up to logarithmic corrections in $\Omega$. The finiteness of $\lambda_\text{min}$ implies that the prethermal fixed point is unstable.

      {\it Dynamical freezing} manifests as an approximate emergent symmetry of $H_0(\lambda)$ in the prethermal regime, for all $\Omega^{-1}\lesssim \lambda \lesssim \lambda_\text{min}$. Consider the emergent symmetry associated with the bare drive component in the Hamiltonian, $H_1(0)=H_{-1}(0)=A{\cal{Q}}$, where ${\cal{Q}}$ is the ``charge" associated with the emergent symmetry, and $A$ is the drive amplitude. The freezing phenomenon occurs in the fast and strong drive limit (i.e. large $A,\Omega$), where the ratio $A/\Omega$ takes discrete special values. At the freezing point, the prethermal Hamiltonian $H_0(\lambda)$ shows an emergent approximate symmetry, characterized by the commutator $[H_0(\lambda),{\cal{Q}}]\sim \mathcal{O}(1/\Omega^2)$, which vanishes in the $\Omega\to\infty$ limit. 
      Therefore, freezing is only approximate at any finite $\Omega$, and importantly, does {\it not} survive when the system flows out of the prethermal regime.

      \item[(c)] When $\lambda\gtrsim\lambda_\text{min}$, the system enters the thermalization regime, where $H_0(\lambda)$ flows to a more non-local Hamiltonian. When the drive frequency $\Omega$ is smaller than a system-size dependent threshold $\Omega_*(L)$, the system eventually flows to a ``thermal" fixed point, where the operator-entanglement entropy is close to that of a random-matrix~\cite{opee1}. Based on our numerical results on small system sizes (see Sec.~\ref{sec:thermalization}), we find that $\Omega_*(L)$ weakly increases with $L$. Whether the threshold diverges in the thermodynamic limit or saturates to a finite value will determine the eventual fate of freezing, which remains an exciting direction to analyze further via new numerical developments in future work.
      From the explicit integration of the flow equations, we find that this complex process occurs through a series of instanton events, where $H_1(\lambda)$ becomes sizable in a narrow window of fRG time, and in the same window a pair of eigenvalues of $H_0(\lambda)$ approach each other. As a result of the instanton events, eigenvalues of $H_0(\lambda)$ will ultimately be pushed into an interval of width $\Omega$, which means $H_0(\lambda\to\infty)$ is now interpreted as a Floquet Hamiltonian. A more detailed technical discussion of these timescales are presented in Sec.~\ref{sec:analytics} and Appendix \ref{app:timescale}. 
 
  \end{itemize}
 
  We end this section by noting that in our numerical computations, we can readily access the ramp-up and prethermal regimes, respectively. In order to access the thermalization regime, we have to restrict ourselves to relatively small drive frequencies due to the limitations associated with numerical floating-point precision.

\begin{table*}[t]
    \centering
    \begin{tabularx}{\linewidth}{||X|X||X|X||}
    \hline\hline
    {\bf fRG time ($\lambda$)} & {\bf Physical Significance} & {\bf Real time ($t$)} & {\bf Physical Significance} \\
    \hline\hline
      Ramp-up  regime: $\mathcal{O}(1)\times \Omega^{-1}$   & RG timescale beyond which the effective Hamiltonian $H_0(\lambda)$ approaches the prethermal fixed point.  &  $t\ll t_\text{th}$ (defined below) & Real timescale where the Floquet-Magnus expansion is a good approximation of the dynamics. \\
      \hline
      Prethermal regime: $\lambda_\text{min}\sim(1/J)\ln(\Omega/J)$   &  RG timescale where $\Fnorm{H_1(\lambda)}$ attains its minimal value after flowing near the prethermal fixed point. & $t_\text{eff}(\lambda)\sim \Fnorm{H_1(\lambda)}^{-1}\Fnorm{\hat{1}}$  & Real timescale upto which $H_0(\lambda)$ serves as a reliable approximation for dynamics.   \\ 
      \hline 
      Thermalization: $\lambda_\text{th}\gtrsim \lambda_\text{min}$ 
      & RG timescale where the system flows away from the prethermal fixed point.  & Thermalization: $t_\text{th}=t_\text{eff}(\lambda_\text{min})\sim(\Omega/J)^{\Omega/J}$ & 
    Real timescale beyond which the prethermal Hamiltonian $H_0(\lambda_\text{min})$ is no longer reliable.  \\
    \hline\hline 
    \end{tabularx}
    \caption{Glossary of flow renormalization group ($\lambda$) and real ($t$) time scales that describe the Floquet dynamics based on the Wegner-Wilson flow framework described in this work.}
    \label{tab:timescales}
\end{table*}

\section{Periodically driven harmonic oscillator}\label{sec:HO}
\begin{figure}[tb]
    \centering
    \includegraphics[width=1.0\linewidth]{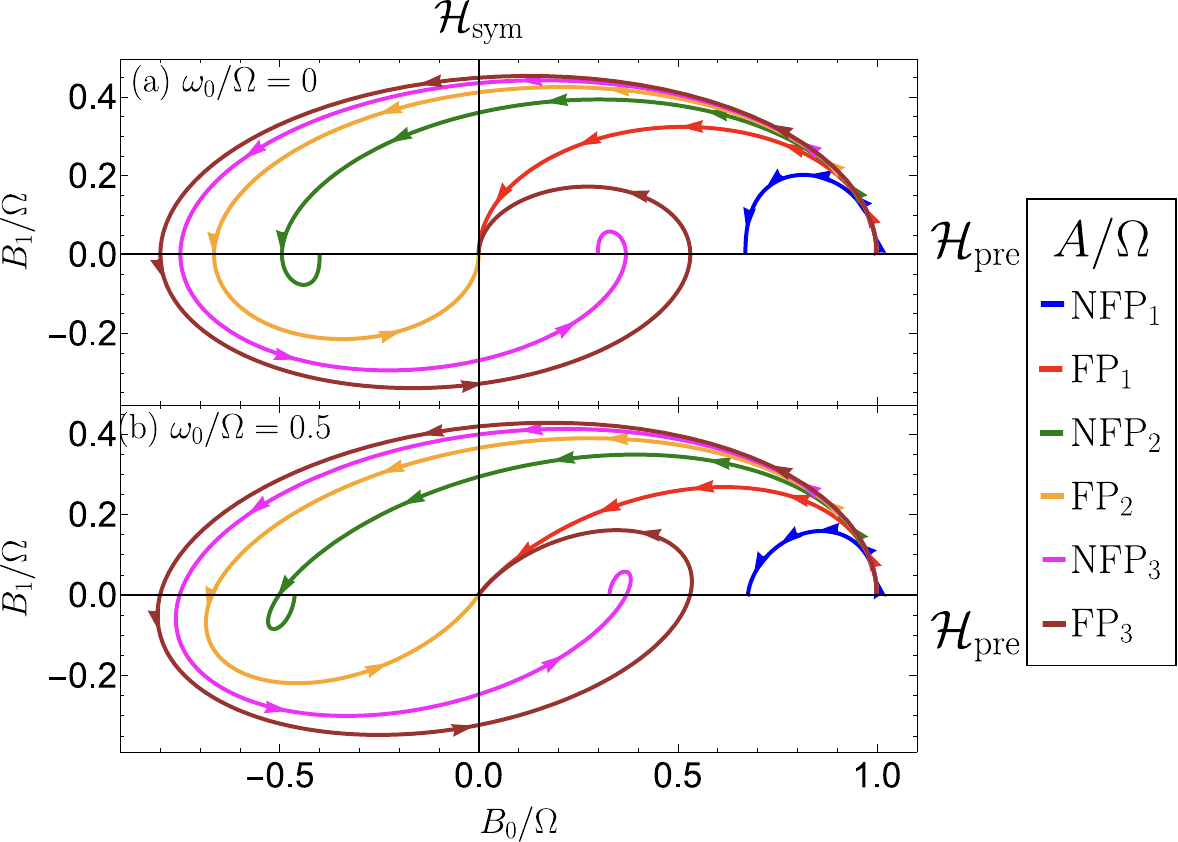}
    \caption{Flow trajectories of a driven harmonic oscillator given by Eq.~\eqref{eq:HO_diff} for $\omega_1=\Omega$, (a) $\omega_0/\Omega=0$, and (b) $\omega_0/\Omega=0.5$ in the $(B_0,B_1)$ plane, depicted for varying $A/\Omega$. The $n-$th freezing point, denoted $\text{FP}_n$, corresponds to distinct values of $A/\Omega$ in (a) and (b), while $\text{NFP}_n$ stands for near-$\text{FP}_n$. $\calH_\text{pre}$ denotes the subspace of prethermal Hamiltonians (locus  $B_1=0$), and $\calH_\text{sym}$ denotes subspace of $\rm{U}(1)$ symmetric Hamiltonians (locus $B_0=0$). Dynamical freezing corresponds to the flow landing exactly on the intersection of $\calH_\text{pre}$ and $\calH_\text{sym}$, as shown by the curves associated with $\text{FP}_1$ (red), $\text{FP}_2$ (orange), and $\text{FP}_3$ (brown).
}\label{fig:HO}
\end{figure}
To illustrate some of the key ideas used to analyze the flow towards dynamical freezing using the above method, we consider a simple (but illustrative) example of a driven non-interacting Hamiltonian,
\begin{equation} \label{eq:HO}
    H(t)=\omega_0 \hat{a}^\dagger\hat{a}+\omega_1(\hat{a}+\hat{a}^\dagger)+ A\hat{a}^\dagger\hat{a}\cos(\Omega t)\,,
\end{equation}
were $\hat{a}^\dagger (\hat{a})$ represents a bosonic creation (annihilation) operator, and $A,~\Omega$ represent the drive amplitude and frequency, respectively. Rewriting the above in terms of the decomposition in Eq.~\eqref{eq:Hbare},
\begin{subequations}
  \begin{eqnarray}\label{}
    H_0(\lambda)&=&A_0(\lambda)\hat{a}^\dagger \hat{a}+B_0(\lambda)\hat{a}^\dagger+B_0^*(\lambda)\hat{a}\,, \\
    H_1(\lambda)&=& A_1(\lambda)\hat{a}^\dagger \hat{a}+B_1(\lambda)\hat{a}^\dagger+C_1^*(\lambda) \hat{a}\,.
  \end{eqnarray}
\end{subequations}
Here $A_0(\lambda)$ and $A_1(\lambda)$ are real functions, while $B_0(\lambda),B_1(\lambda),C_1(\lambda)$ can be complex, and we have dropped the parts proportional to the identity operator. The two flow equations Eq.~\eqref{eq:FlowH0} and \eqref{eq:FlowH1} become
  \begin{subequations}\label{eq:HO_diff}
    \begin{eqnarray}
    \frac{\partial A_0}{\partial \lambda} &=& 0\,, \\
    \frac{\partial A_1}{\partial \lambda} &=& -\Omega A_1 \,, \\
    \frac{\partial B_0}{\partial \lambda} &=& 2 A_1(C_1-B_1)\,, \\
    \frac{\partial B_1}{\partial \lambda} &=& -\Omega B_1-A_0 B_1+A_1 B_0 \,, \\
    \frac{\partial C_1}{\partial \lambda} &=& -\Omega C_1 +A_0 C_1-A_1 B_0 \,,
  \end{eqnarray}
  \end{subequations}
  and the initial conditions at $\lambda=0$ are $A_0=\omega_0,A_1=A/2,B_0=\omega_1,B_1=C_1=0$.
The flow equations reduce to first-order differential equations for $A_{0,1}$, $B_{0,1}$, and $C_1$, respectively. Solutions for  $A_0(\lambda)=A_0$ (constant) and $A_1(\lambda)=(A/2)e^{-\Omega\lambda}$, while others can be solved numerically. Fig.~\ref{fig:HO} depicts flow trajectories in the $(B_0,B_1)$ plane for varying ratios of $A/\Omega$ and $\omega_0/\Omega$, respectively. Due to integrability of Eq.~\eqref{eq:HO}, the system directly flows to the prethermal fixed point, with $\lambda_\text{min}=\infty$ and $\Fnorm{H_1(\lambda_\text{min})}=0$.

  \begin{figure*}[htb!]
    \centering
    \includegraphics[width=1.0\linewidth]{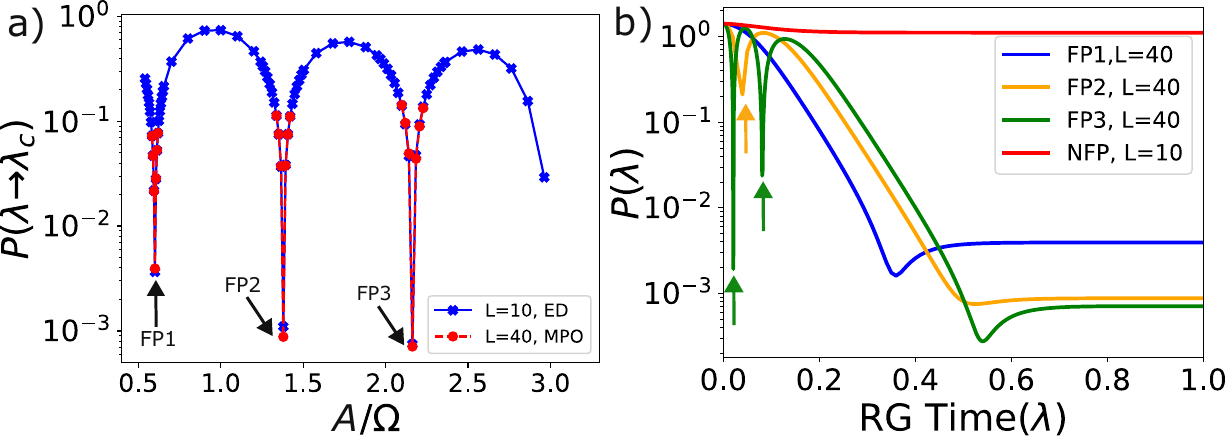}
    \caption{(a) $P(\lambda)$ defined in Eq.~\eqref{commH0Xnorm} as a function of $A/\Omega$ for $\lambda\rightarrow\lambda_c$. The sharp drop in $P(\lambda_c)$ marks the freezing points, FP$_1$, FP$_2$, and FP$_3$, in the considered range of $A/\Omega$.  
    The results for $P(\lambda)$ are obtained for system size \( L=10 \) using exact diagonalization, and for \( L=40 \) from MPO calculations. (b) $P(\lambda)$ as a function of fRG time, (\( \lambda \)), for the different freezing points shows a saturation to the plateau at ``late" times. Results for all freezing points are shown for \( L=40 \) from MPO computations, and for non freezing point (NFP) for $L=10$ using ED. Other parameters: $J=1$, $J_{2}=0.2$, $\Omega=10$. We show the results for $P(\lambda)$ for $L=10$ until $\lambda=10$ in the Appendix~\ref{app:Supp}.   }\label{fig:freezing}
\end{figure*}

  In this example, dynamical freezing is associated with an emergent $\rU(1)$ symmetry generated by ${\cal{Q}}=\hat{a}^\dagger \hat{a}$, with the distance between $H_0(\lambda)$ and $\calH_\text{sym}$ --- the space of U(1)-symmetric Hamiltonians --- measured by $|B_0(\lambda)|$.  The flow in Fig.~\ref{fig:HO} follows a spiral trajectory in the $(B_0,B_1)$ plane. Since at the end of the flow, the oscillatory part of the Hamiltonian is completely removed, the flow will end on the $B_1=0$ line, the space of prethermal Hamiltonians $\calH_\text{pre}$. For a non-freezing point, the flow ends on $\calH_\text{sym}$ with $B_0(\lambda\to\infty)\neq 0$. This corresponds to the presence of a $\rm{U}(1)$ breaking term in the renormalized Hamiltonian. In contrast, freezing occurs for $B_0(\lambda\to\infty)=0$ (equivalently, $[H_0(\lambda\to\infty),{\cal{Q}}]=0$),
  linked to certain initial values of $A/\Omega$, which generally depend on $\omega_0$. In the limit $\omega_0/\Omega\to 0$, these values align with the Magnus expansion, namely, the zeros of the Bessel function $J_0(...)$ (see Sec.~\ref{sec:analytics} and Appendix.~\ref{app:magnus} for more details). Since $H_0(\lambda\to\infty)$ depends on $A/\Omega$ continuously, the order of the freezing point can be related to the geometry of the flow trajectory.  In particular, every time $H_0(\lambda)$ crosses the axis $B_0=0$, the order of the freezing point increases by one; see Fig.~\ref{fig:HO}.

  As a caveat, some of the features of the simple example shown above \emph{do not} generalize to a generic many-body Hamiltonian. First, here the system is fully integrable and does not ``thermalize" in the conventional sense, meaning that the flow only explores the ramp-up and prethermal regimes, respectively. Second, since there is only one set of terms ($\hat{a},~\hat{a}^\dagger$) that violates the $\rm{U(1)}$ symmetry, we can realize freezing for any oscillator frequency $\omega_0/\Omega$ by tuning the parameter $A/\Omega$. In a generic many-body system, there can be multiple symmetry violating operators in $H_0(\lambda)$, and $[H_0(\lambda),\calQ]$ need not vanish by tuning a single parameter $A/\Omega$. However, as we will show later in Sec.~\ref{sec:spinchain}, the minimum value decays as a power-law in $1/\Omega$. Moreover, many of the qualitative features discovered here via the flow trajectories generalize to varying degrees in the more generic setting, as we will highlight below.

\section{Floquet spin chain: numerical results}\label{sec:spinchain}

We now turn to the main example of interest in this paper, and demonstrate that the key features discussed above can be generalized to a many-body setting using the same fRG. We have analyzed the flow trajectories for a one-dimensional, non-integrable Ising spin chain with a Floquet cosine-wave drive coupled to the transverse field. The results are sufficiently universal, even in the presence of additional perturbations to the Hamiltonian, that we believe our results are representative of the general phenomenology introduced above. The Hamiltonian is given by,
\begin{equation}
H(t)=\sum_i\left[-J S_i^z S_{i+1}^z-J_2 S_i^z S_{i+2}^z+2A S_i^x \cos(\Omega t)\right] \,,
\end{equation}
where the second-neighbor coupling $J_2$ is responsible for the non-integrable nature; for $J_2=0$, freezing in the integrable (free) Hamiltonian has been analyzed previously~\cite{ADas2010}. Note that in the absence of the drive term, the above Hamiltonian is entirely classical. Hence we have also analyzed the flow RG in the vicinity of freezing points as a function of a {\it static} transverse magnetic field in the $x-$direction, $B_x$, for the modified Hamiltonian,
\beq
H(t)\rightarrow H(t) + \sum_i B_x S_i^x.
\eeq

We can split the above Hamiltonian into two parts, 
\begin{subequations}
 \beq\label{Eq:SpinH0}
    H_0(0)&=& \sum_{i}\left[-J S^z_i S^z_{i+1}-J_2 S^z_{i}S^z_{i+2} + B_x S_i^x\right]\,,\\
    H_1(0)&=&A \sum_{i} S^x_i\,,\label{Eq:SpinH1}
  \eeq
\end{subequations}
which represent the initial condition for the flow associated with $H_{0}(\lambda)$ and $H_{\pm 1}(\lambda)$, respectively. Our numerical results will be presented in units where $J=1$. Since the phenomenology remains qualitatively similar with the inclusion of $B_x$, we first begin by setting $B_x=0$, and then study the problem as a function of increasing (but static) $B_x$. Nevertheless, there are interesting quantitative modifications to the freezing phenomenology with the inclusion of $B_x\neq0$, which we will discuss briefly in Sec.~\ref{sec:transverse} below, and elaborate further in greater detail in Appendix~\ref{app:Supp}

At the dynamically frozen point, we expect the operator $\mathcal{O}=\sum_i S_i^x(\equiv{\cal{Q}})$ to exhibit an approximate emergent conservation law. One of the principle quantities of interest will be,
\begin{equation}\label{commH0Xnorm}
    P(\lambda)=\frac{\left\Vert \left[H_0(\lambda),{\cal{O}}\right]\right\Vert}{\left\Vert H_0(\lambda)\right\Vert}\,,
\end{equation}
where $\Fnorm{M}=\sqrt{\text{Tr}(MM^\dagger)}$ is the Frobenius norm. A small $P(\lambda)$ implies approximate emergent symmetry generated by $\mathcal{O}$. To compare across different system size, we have normalized $P(\lambda)$ to be an intensive quantity.
It is also useful to define the intensive ratio of the norms of the driven part of the Hamiltonian with respect to the static part,
\begin{equation}\label{H1norm}
    Q(\lambda)=\frac{\left\Vert H_{1}(\lambda) \right\Vert}{\left\Vert H_{0}(\lambda) \right\Vert}\,,
\end{equation}
which quantifies the extent to which the oscillatory component has been removed during the flow. As discussed in the discussion surrounding Eq.~\eqref{eq:teff}, a smaller $Q(\lambda)$ implies that $H_0(\lambda)$ is a better approximation for the effective Floquet Hamiltonian upto later times.

\begin{figure*}[htb!]
    \centering
    \includegraphics[width=1.0\linewidth]{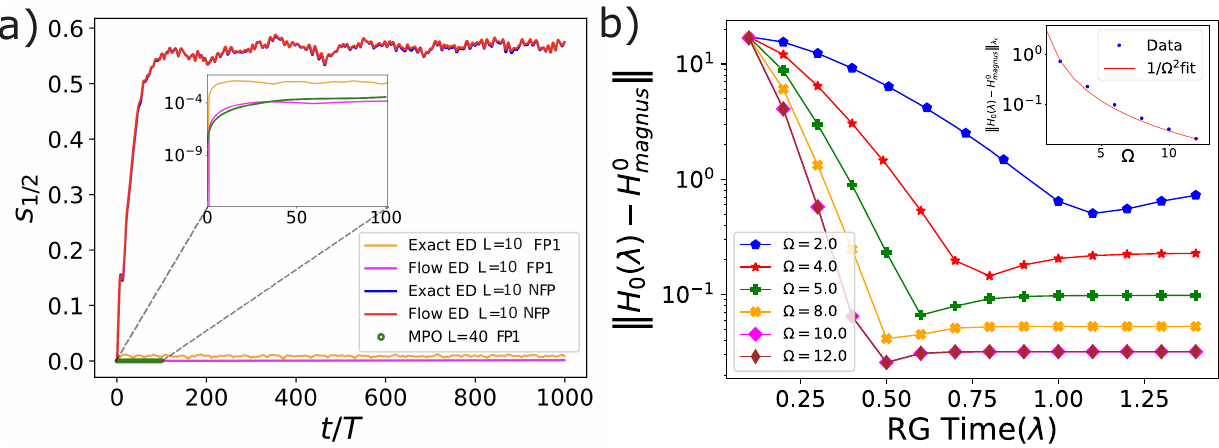}
    \caption{(a) Half chain entanglement entropy (EE) density, [i.e. $s_{1/2}=S_{1/2}/(L/2))$] plotted stroboscopically as a function of real time starting from a fully polarized state $\ket{\psi}=\ket{x;+}^{\otimes L}$. The results are shown for $L$=10 (using both exact and flow ED), and $L=40$ (using Flow MPO/MPS), at and away from freezing. Inset: Early time evolution of $s_{1/2}$ at FP$_1$. Other parameters are set to $J=1$, $J_{2}=0.2$, and $\Omega=10$. (b) Frobenius norm of $H_{0}(\lambda)-H^{0}_\text{magnus}$  as a function of fRG time for varying $\Omega$. $H^0_\text{magnus}$ is computed analytically in Appendix~\ref{app:Magnus_spinchain}. Inset: Saturation value at late fRG time falls off as $1/\Omega^2$ at FP$_1$ ($L=10$).}\label{fig:EEmag}
\end{figure*}

\subsection{Approach to prethermal regime \\at freezing point}

To diagnose dynamical freezing numerically, we will evolve Eqs.~\eqref{eq:Flow} (i.e. let them flow) to a fRG time $\lambda=\lambda_c$, such that  $Q(\lambda_c)\ll 1$ and $P(\lambda)$ exhibits a $\lambda-$independent plateau, as introduced in regime (b) in Sec.~\ref{sec:flow}; see also Table~\ref{tab:timescales}. In other words, $\lambda_c$ is then in the initial prethermal regime.  In this subsection, we choose $\lambda_c\lesssim\lambda_\text{min}$, the fRG time where $H_1(\lambda)$ minimizes. For all of the numerical computations in this section, we find that a $\lambda_{c}=1\times J^{-1}$ satisfies these numerical criteria. In general, we find that $\lambda_c$ should be set to the inverse intrinsic single-particle energy scales associated with the local thermalization dynamics. 

In Fig.~\ref{fig:freezing}(a), we show explicit results for $P(\lambda)$ for $B_x=0$ at and away from the freezing points; clearly, this requires us to identify the location for the freezing points. Thus, we first performed a scan of $P(\lambda=\lambda_c)$ over a fine interval of $A/\Omega$ (including for a $L=40$ chain using MPO), which clearly reveals the sharp features associated with freezing. These are encoded in the striking drop of $P(\lambda_c)$ at specific values of $A/\Omega$. Note that $\lambda_c$ is chosen such that $P(\lambda)$ is already in a plateau-like regime, as explained above, and shown in Fig.~\ref{fig:freezing}(b). Interestingly, there is an ``asymptotic" plateau associated with $P(\lambda)$ as a function $\lambda$, both at and away from freezing (We show in Appendix~\ref{app:Supp} that the plateau feature at $\Omega=10J$ can extend to much longer fRG time). However, the flow towards the plateau at the first, second, and third freezing points, are accompanied by a rich structure, reminiscent of some aspects of the flow trajectories for the driven harmonic oscillator in Fig.~\ref{fig:HO}, which we shall describe in more detail below. We note here that the data in Fig.~\ref{fig:freezing}(a) are obtained for a fixed $\Omega$ and varying $A$. The numerical results remain identical, if instead $A$ is fixed and $\Omega$ is varied (see Appendix \ref{app:Supp}). However, the only requirement remains that both $A$ and $\Omega$ are large compared to the single particle energy scales in $H(t)$, with $A/\Omega\sim O(1)$ for FP$_n$ with $n\sim O(1)$.

The locations of the freezing points are roughly consistent with the expectations based on Floquet-Magnus expansion, namely $J_0(A/4\Omega)=0$, where $J_0$ is the Bessel function (see Appendix~\ref{app:Magnus_spinchain} for derivation). As we demonstrate in Appendix~\ref{app:Supp}, the position of the freezing point is renormalized away from the leading-order Magnus result, with a stronger renormalization for smaller $\Omega$.

The flow initially shows a ramp-up behavior where $H_0(\lambda)$ (and relatedly, $P(\lambda)$) varies rapidly with the fRG time, $\lambda$. This is followed by the prethermal regime where $H_0(\lambda)$ flows slowly with $\lambda$, and as a result, $P(\lambda)$ reaches a plateau. As noted before, the hallmark of ``dynamical freezing" is the strong reduction in the value associated with the plateau in $P(\lambda)$. It is worth highlighting that this plateau value at freezing, while at least an order of magnitude smaller than the initial value $P(0)$, does not vanish exactly. This leads to only an approximate conservation of $\mathcal{O}$. In Sec.~\ref{sec:transverse} we will return to this point when $B_x\neq0$.

We note two additional features associated with $P(\lambda)$ at freezing. First, they develop strong ``dips" in the ramp-up regime (marked by arrows in Fig.~\ref{fig:freezing}b), and the number of such dips is given by $(n-1)$, where $n$ represents the order of freezing point. This is reminiscent of the driven harmonic oscillator example in Fig.~\ref{fig:HO}, where with varying $A/\Omega$, the flow trajectory can cross the $\calH_\text{sym}$ plane, associated with the subspace of symmetric Hamiltonians. In the many-body situation, the crossing is not exact. Nevertheless associated with each such near crossing, the distance (in the space of Hamiltonians) between the effective Hamiltonian and $\calH_\text{sym}$, which is given by $P(\lambda)$ develops a minimum. This corresponds to the dips. Second, for higher drive amplitudes $A/\Omega$, the flow takes a longer fRG time to reach the plateau. This can be intuitively understood as follows: the flow effectively integrates away the oscillatory part, $H_1(\lambda)$, at a rate proportional to the drive frequency $\Omega$, such that a stronger drive spends a longer time in the ramp-up regime. This feature is also consistent with our analytical results in Sec.~\ref{sec:analytics}.

It is worth emphasizing that the finite value of $P(\lambda)$ at freezing is in sharp contrast to the harmonic oscillator example in Fig.~\ref{fig:HO} presented earlier, where $P(\lambda)$ vanishes exactly. The difference can be traced back to our concluding comments in Sec.~\ref{sec:HO}, where we pointed out that in a many-body system, a variety of operators that violate the emergent conservation can be generated along the flow, which cannot all be suppressed by fine-tuning a single parameter, $A/\Omega$. However, these additional symmetry violating operators can be universally suppressed in the high frequency $\Omega\to\infty$ limit. We find $P(\lambda_c)$ scales with the frequency as $1/\Omega^2$ at the freezing points (see Appendix~\ref{app:Supp} for more detail).

Let us end by relating our findings based on the flow formalism to the real-time dynamics of the driven system. As discussed in Sec.~\ref{sec:flow}, the static part of the Hamiltonian $H_0(\lambda)$ is to be interpreted as an approximation for the actual Floquet Hamiltonian, $H_F$, with an error bounded by $\Fnorm{H_1}$. As 
discussed in Appendix~\ref{app:Supp}, $\Fnorm{H_1(\lambda_c)}$ decays exponentially with $\Omega$, implying that $H_0(\lambda_c)$ should be an excellent approximation to $H_F$ in the large$-\Omega$ limit. To test this, we have computed the growth of the half-chain entanglement entropy density of the system, $s_{1/2}=S_{1/2}/(L/2)$, starting from the polarized state $\ket{\psi}=\ket{x;+}^{\otimes L}$, using two different methods (see Appendix \ref{app:numericaldetails} for additional numerical details): 
\begin{itemize}
    \item[(a)] We use the exact real-time evolution based on exact diagonalization to evaluate $s_{1/2}$.
    \item[(b)] We treat the problem as an effective static system, but with Hamiltonian $H_0(\lambda_c)$, obtained from the solutions to the flow-equations using exact diagonalization, as well as MPO-based evolution using Time-Dependent-Variational Principle (TDVP) \cite{tdvp1,tdvp2,fishman2022itensor,Itensor2,Yang_2020,MPSevolution2}, respectively. 
\end{itemize} 
The results for $s_{1/2}$ obtained using both sets of methods are plotted in Fig.~\ref{fig:EEmag}(a), which show excellent numerical agreement at and away from freezing. This provides further evidence for our Floquet flow renormalization being able to capture the essential universal aspects of the problem, including the growth of entanglement entropy in real-time. We also note that $H_0(\lambda)$ in the prethermal regime accurately captures the real-time dynamics up to an exponentially long time. It is worth noting that the growth of entanglement entropy in Fig.~\ref{fig:EEmag}(a) shows clear differences between freezing points (FP1) and non-freezing points (NFP). Clearly at early to intermediate times, the freezing phenomenology is associated with much slower dynamics; however, this does not imply that this difference would survive up to much later times. The growth of entanglement is exponentially slow in the drive frequency as well, which can be seen quite clearly by focusing on smaller system sizes \cite{supp}.

Finally, we directly compare the ``distance" between $H_0(\lambda)$ and $H^0_\text{magnus}$ (quantified by Frobenius norm) in Fig.~\ref{fig:EEmag}(b), the leading order result of the Floquet-Magnus expansion (see Appendix~\ref{app:Magnus_spinchain} for detail), which is expected to capture the prethermal dynamics in the $\Omega\to\infty$ limit. Indeed, we see that this distance decreases along the flow trajectory, $\lambda$, and reaches a plateau in the prethermal regime; the plateau value decays as $1/\Omega^2$ in the high frequency limit, which is consistent with our analytical result in Sec.~\ref{sec:analytics}. Furthermore, $H_0(\lambda_c)$ is approximately isospectral to the actual Floquet Hamiltonian $H_F$ \cite{supp}.

\subsection{Effect of a static transverse magnetic field}
\label{sec:transverse}
\begin{figure}[htb]
    \centering
    \includegraphics[width=1\linewidth]{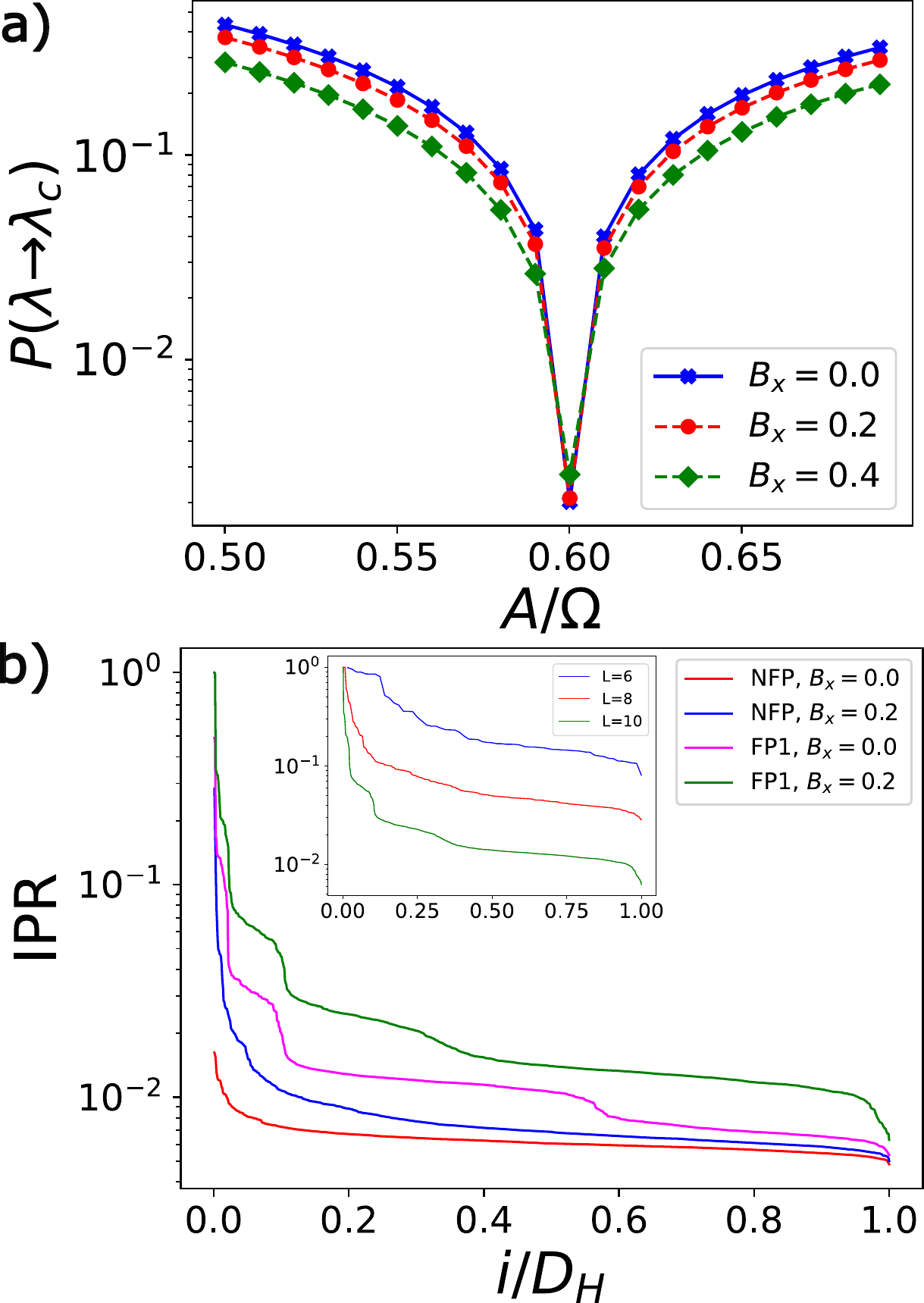}
    \caption{(a) $P(\lambda \to \lambda_{c})$ obtained from Eq.~\eqref{commH0Xnorm} as a function of $(A/\Omega)$ near FP$_1$ with increasing $B_{x}$ for $L=10$. All other parameters are same as in Fig.~\ref{fig:freezing}. (b) IPR defined via Eq.~\eqref{ipr} for eigenstates of $H_{0}(\lambda_c)$ in the $\hat{S}_x$ basis plotted in decreasing order for $L=10$, comparing the freezing point (FP1) and the nonfreezing point (NFP), both with and without the magnetic field term in the $x$ direction. Inset: IPR vs $i/D_H$ for increasing system size for FP$_1$ with $B_x\neq0$. Other parameters are: $J=1$, $J_2=0.5$, $\Omega=5$, $\lambda_c=5.0$.}\label{fig:transverseBfield}
\end{figure}

As noted previously, $H_0$ is purely classical when $B_x = 0$; see Eq.~\eqref{Eq:SpinH0}. Here, we analyze the effect of introducing an additional static transverse field along the $x$ direction. Specifically, we study the norm of $P(\lambda)$ as in Eq.~\eqref{commH0Xnorm} near the dynamical freezing point. In Fig.~\ref{fig:transverseBfield}(a), we present results for $P(\lambda \to \lambda_{c})$ as a function of $(A/\Omega)$ near FP$_1$ for increasing values of $B_x$. These results demonstrate that the location of the freezing point associated with the first (prethermal) fixed point is unaffected by increasing $B_x$, as indicated by the nearly identical locations of the minimum in $P(\lambda_c)$ for all $B_x$. 

However, the robustness (i.e. ``quality") of freezing depends on the value of $B_x$, as we now demonstrate by investigating the Floquet eigenstates. To analyze how the nonzero transverse field affects the Floquet eigenstates, we compute their localization properties in the many-body Hilbert space. Specifically, we calculate the inverse participation ratio (IPR) of a Floquet eigenstate, $\ket{m}$, in the eigenbasis ($\ket{i_x}$) of $\hat{S}_x$, 
\beq
\tn{IPR} = \sum_{i_x=1}^{D_H} |\langle i_x | m \rangle|^4,
\label{ipr}
\eeq 
where $D_H=2^L$ is the dimension associated with the many-body Hilbert space. When $\ket{m}$, is localized over only a few (i.e. $O(1)$) of the eigenstates $\ket{i_x}$, we expect $\tn{IPR}\sim O(1)$. In contrast, when $\ket{m}$ is delocalized 
over all of the eigenstates in the eigenbasis of any local operator (e.g. $\hat{S}_x$), the IPR falls off rapidly with increasing system-size. In the present setting of a driven many-body system, where the drive couples to $S_i^x$, unbounded heating would lead to increasing delocalization and decreasing $\tn{IPR}$. In Fig.~\ref{fig:transverseBfield}(b), we observe that at the freezing point, the IPR increases and approaches an $O(1)$ value with increasing $B_x$, indicating enhanced localization in the many-body Hilbert space. Interestingly, by analyzing the IPR for the Floquet eigenstates of $H_{0}(\lambda_c)$ in the $x$ basis both at freezing (FP1) and away from freezing (NFP), as well as with and without a finite $B_x$, we find a higher tendency towards localization when $B_x\neq0$. However, the Floquet eigenstates also develop an increased tendency towards delocalization in the eigenbasis of any local operator, as evaluated based on the IPR for the Floquet eigenstates of $H_{0}(\lambda_c)$. At the freezing point, the effective Floquet Hamiltonian to leading order is proportional to the $YY + ZZ$ terms; see Appendix \ref{app:Magnus_spinchain}. This structure gives rise to an emergent $\mathbb{Z}_2$ symmetry, resulting in degenerate Floquet eigenstates (equivalently, eigenstates of $H_0(\lambda_c)$), and consequently, the expectation values of $S_x$ are nearly zero for all states. Introducing a small nonzero $B_x$ lifts the degeneracy, making the Floquet eigenstates to align with the eigenstates of $S_x$. We have further analyzed the effect of $B_x$ on the expectation values of the magnetization in Appendix \ref{app:Supp}.

\subsection{Thermalization via instantons}
\begin{figure}[t]
    \centering
    \includegraphics[width=1.0\linewidth]{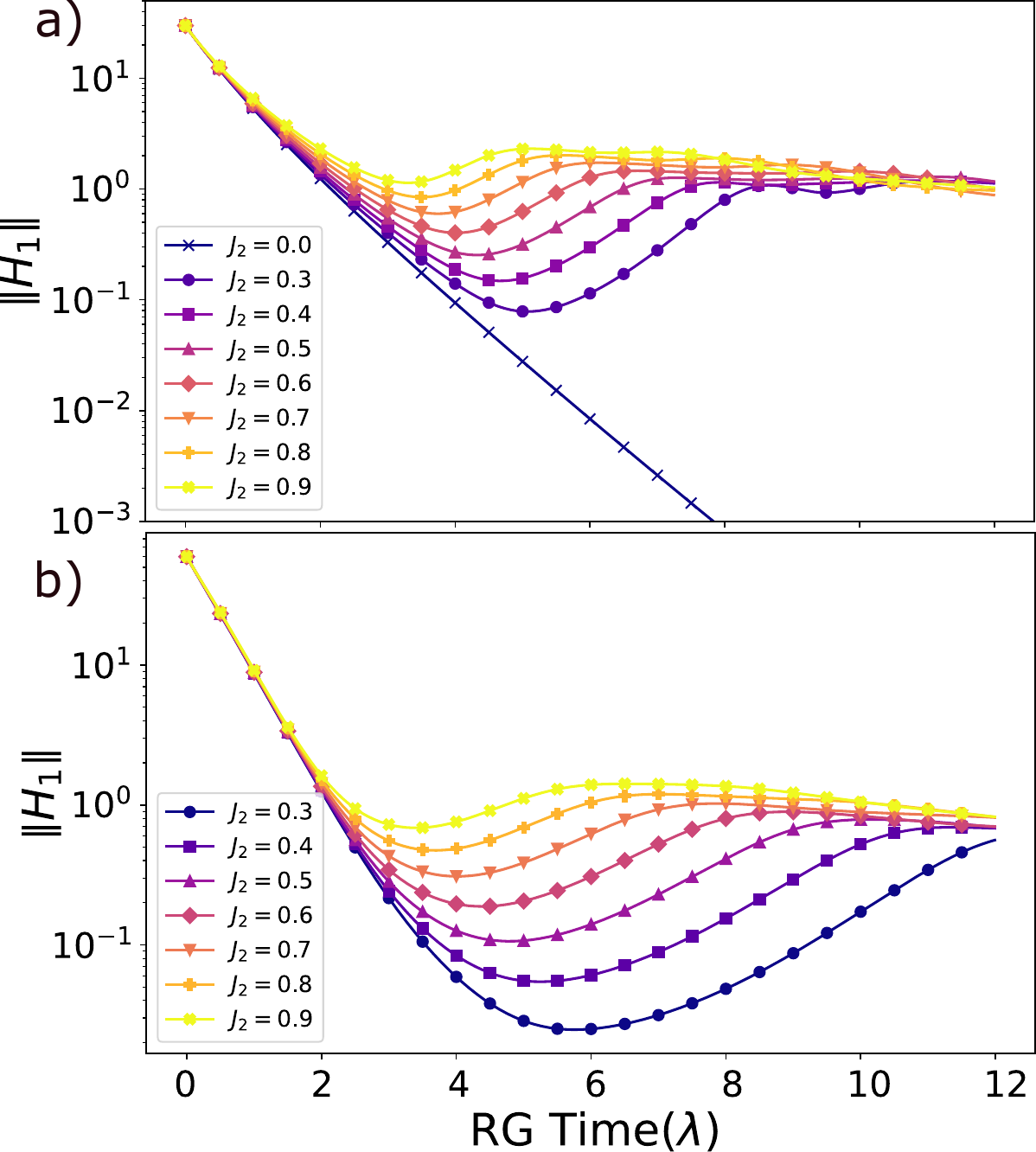}
    \caption{The Frobenius-norm of (a, b) $H_{1}$ as a function of fRG time for increasing $J_{2}$ for $L=10$. (a) Away from freezing, and for \( J_{2}=0 \), the system flows toward an integrable fixed point. For \( J_{2} \neq 0 \), at small fRG times the system first flows toward a prethermal fixed point, and eventually flows to a thermal fixed point. (b) At freezing, the system flows toward a prethermal fixed point for small and intermediate fRG times, controlled by the emergent approximate integrability, with the subsequent flow to the thermal fixed point delayed relative to (a).  Other parameters are set to: $\Omega=2J$, $J=1$, $A/\Omega=0.3$ away from freezing, $A/\Omega=0.601$ at FP$_1$. }\label{fig:L40rf}
\end{figure}

Let us now address the flow away from the prethermal fixed point. For that matter, we study $\Fnorm{H_1(\lambda)}$ for different values of the next-nearest-neighbor coupling $J_2$, at and away from a freezing point, up to intermediate fRG time, $\lambda_c$. In this subsection, we will set $\lambda_c\gg \lambda_\text{min}$.
We note that $J_2\neq 0$ breaks free-fermion integrability of the model. Therefore, we expect that $H_1(\lambda)$ ceases to decay, and shows an upturn at intermediate fRG times when $J_2\neq 0$. Our access to this regime is partly constrained by two factors: First, the initial exponential decay of $H_1(\lambda)$ is limited by numerical floating-point precision. The prethermal cutoff fRG time $\lambda_\text{min}$ and the drive frequency $\Omega$ are limited by $\Omega \lambda_\text{min}<\ln(1/\epsilon_\text{fp})$, where $\epsilon_\text{fp}\sim 10^{-16}$ is the floating-point precision. Beyond that, $H_1(\lambda)$ will decay below floating-point precision before reaching the upturn. Second, the growth of $\Fnorm{H_1(\lambda)}$ is associated with a non-trivial spreading of operators in space, causing the bond dimension of the MPO to grow rapidly. Due to these constraints, we choose to study this regime using the ED method with a smaller value of $\Omega$, as shown in Fig.~\ref{fig:L40rf}. The latter issue can be addressed using more sophisticated MPO compression in future work \cite{DEParker2020}, but we expect that the qualitative considerations are likely to be similar to what we describe below.  

As noted before, for a small fRG time $\lambda$ in the ramp-up regime, $H_1(\lambda)$ decays exponentially with $\lambda$, as shown in Fig.~\ref{fig:L40rf}. However, for intermediate values of $\lambda$, $\Fnorm{H_1(\lambda)}$ reaches a minimum and starts to grow at $\lambda=\lambda_\text{min}$. Concurrently, the static component $H_0(\lambda)$ is no longer in the plateau regime, and starts to flow as well. Using Eq.~\eqref{eq:teff}, we can interpret the minimum of $\Fnorm{H_1(\lambda)}$ as the termination of the prethermal regime, and the associated time scale $t_\text{eff}\sim\Fnorm{\hat{1}}/\Fnorm{H_1(\lambda_\text{min})}$ marks the maximal (real) time of validity of the prethermal Hamiltonian $H_0(\lambda_\text{min})$. In the fRG language, this implies that the prethermal fixed point is unstable in generic non-integrable systems, and the flow turns away from the fixed point at $\lambda=\lambda_\text{min}$.

Clearly, as the system flows away from the prethermal fixed point, we expect the emergent symmetry associated with dynamical freezing to also be lost. However, we point out that dynamical freezing does provide additional mitigation against thermalization. As we see in Fig.~\ref{fig:L40rf}, for a many-body system driven at the freezing point, the minimum fRG time $\lambda_\text{min}$ of $\Fnorm{H_1(\lambda)}$ is indeed {\it delayed} compared to the same system away from freezing. At the same time, $\Fnorm{H_1(\lambda_\text{min})}$ is smaller, and $H_0(\lambda)$ shows a longer plateau-like interval (See \cite{supp} for the corresponding figure of $H_0(\lambda)$). This implies that the dynamically frozen regime (for the cosine-wave drive) thermalizes slower compared to a generic driven (unfrozen) system, in the sense of both fRG time and real time.

\begin{figure}[t]
    \centering
    \includegraphics[width=1.0\linewidth]{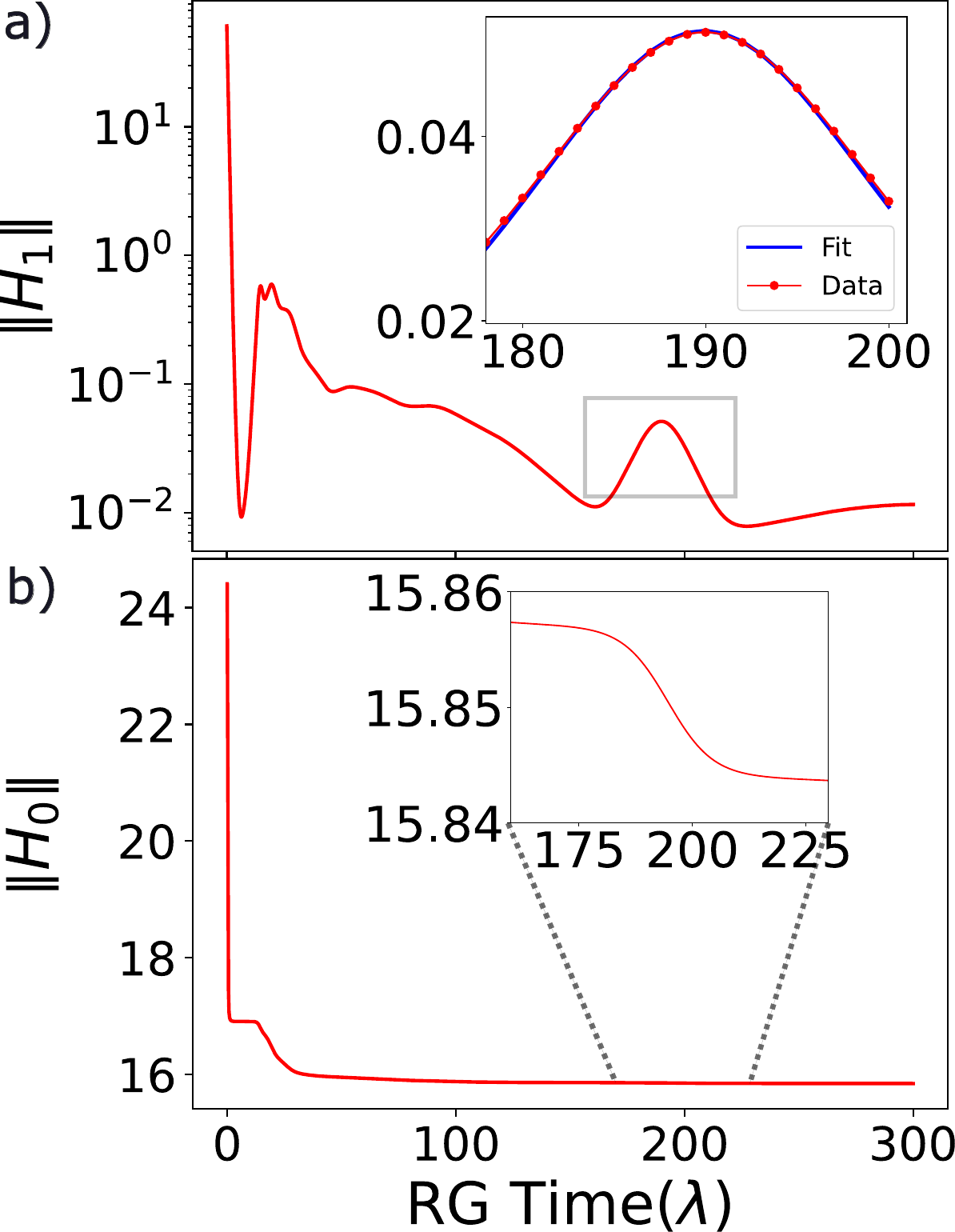}
    \caption{(a) Frobenius norm of $H_{1}(\lambda)$ as a function of fRG time for $L=10$. Inset: isolated peak at late fRG time along with theoretical fit [$(a/2){\rm sech}(a(\lambda-b))$]. Parameters: $J=1$, $J_{2}=0.2$, $\Omega=2.0$ at FP$_1$. (b) Frobenius norm of $H_{0}(\lambda)$, with a step-like feature correlated with the peak in $H_{1}(\lambda)$ in (a). Additional results for $L=6$ are shown in Appendix~\ref{app:Supp}.}\label{fig:H1nL10}
\end{figure}

Next, we address the flow dynamics beyond the prethermal fixed point. We run the flow equation Eq.~\eqref{eq:Flow} using the ED method up to a fRG time $\lambda\gg\lambda_\text{min}$, and the result is shown in Fig.~\ref{fig:H1nL10}. After the minimum associated with the termination of the prethermal fixed point, $\Fnorm{H_1(\lambda)}$ shows additional peaks, separated by regions of exponential decay.  
To interpret the behavior of $\Fnorm{H_1(\lambda)}$, we generalize the lesson we learn from the prethermal fixed point: The flow equation Eq.~\eqref{eq:FlowH0} shows that the flow rate of $H_0(\lambda)$ is determined by the magnitude of $H_1(\lambda)$.  When $\Fnorm{H_1(\lambda)}$ decays exponentially, it implies that the system is close to an unstable fixed point and that the flow of $H_0(\lambda)$ runs slowly. In contrast, when $\Fnorm{H_1(\lambda)}$ grows and attains a peak, it is interpreted as a rapid flow from one fixed point to the other. As we show in Fig.~\ref{fig:H1nL10}, when $H_1(\lambda)$ shows a peak, there is a concurrent step-like feature in $H_0(\lambda)$. Therefore, the trajectory at large fRG times can be interpreted as the flow between a series of unstable fixed-points. In particular, most of the fRG time is spent near a fixed point, and the transition between fixed-points only takes a short interval of fRG time. Therefore, it is appropriate to associate these transitions with instantons \cite{weinberg1995quantum}.

As we will discuss in Sec.~\ref{sec:analytics}, an instanton event is associated with the reorganization of the spectrum of $H_0(\lambda)$. When $\lambda\gg \lambda_\text{min}$, most instantons are elementary in the sense that they only involve a pair of eigenvalues of $H_0(\lambda)$ and a single matrix element of $H_1(\lambda)$, which is amenable to an analytical description. As a result, $\Fnorm{H_1(\lambda)}$ shows a simple `single-peak' structure that agrees with the analytical solution in Eq.~\eqref{eq:nonlinear_solution} below. However, we note there are also instances where multiple instantons are close to each other, and there can be nontrivial interactions between them. As shown in Fig.~\ref{fig:H1nL10}, the first growth regime of $\Fnorm{H_1(\lambda)}$ after $\lambda=\lambda_\text{min}$ is one of such examples, where two instantons are close to one another resulting in a `double-peak' feature in $\Fnorm{H_1(\lambda)}$. In Appendix~\ref{app:Supp}, we present results for a smaller system size ($L=6$), where we were able to simulate the flow for significantly longer fRG times and access more instanton events. We found examples of well-isolated instantons, which continue to be well described by the theoretical results in Sec.~\ref{sec:analytics}. Finally, we remark that the analytical treatment in Sec.~\ref{sec:analytics} relates the width of the instanton with its height, but the exact location and the height of the instanton is determined by the microscopic details of the Hamiltonian. Nevertheless, we expect the first instanton to happen at a RG time scale $\lambda$ that is parametrically comparable to $\lambda_\text{min}$.

\section{Numerical diagnosis of thermalization}\label{sec:thermalization}

Our analysis thus far has established the emergence of prethermal regimes and instanton-mediated dynamics near dynamical freezing. However, we have not addressed the fundamentally important question of the extent to which the system thermalizes and loses memory of its initial state in the late-time limit. To address this question, we now present results for two complementary numerical diagnostics: 
(i) the Operator Entanglement Entropy (OPEE) ~\cite{opee1,opee2} associated with $H_0(\lambda)$, the flowing effective Hamiltonian, which probes the build-up of operator non-locality and complexity, and 
(ii) the Diagonal Ensemble (DE) average of local observables, which captures the infinite-time limit of their expectation value after dephasing in the Floquet basis. 
The OPEE serves as a global measure that is sensitive to random-matrix-like behavior, while the DE average tests for the loss of memory of the initial state. 
Together, these quantities allow us to assess the extent of thermalization at and near freezing, and to identify regimes where prethermal behavior persists upto late times.

\begin{figure*}[t]
    \centering
    \includegraphics[width=1.0\linewidth]{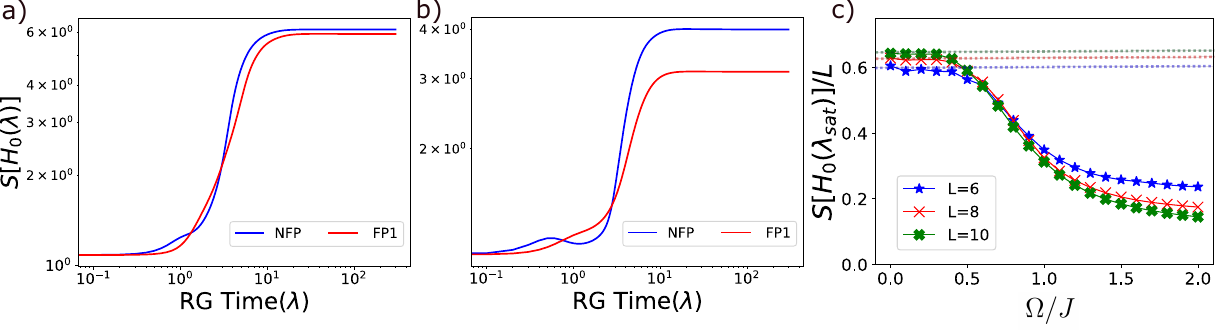}
    \caption{(a) The operator entanglement entropy (OPEE) for the operator $H_{0}(\lambda)$ is plotted as a function of RG time, both at and away from the freezing point ($L=10$, $\Omega/J=0.5$). Both of the curves approach the RMT value. (b) For $\Omega/J=1.0$, at the freezing point, the OPEE is significantly suppressed compared to the non-freezing point. (c) The saturation value of the OPEE is shown as a function of the driving frequency at the first freezing point, across different system sizes. As the frequency increases, the OPEE saturation value approaches sub-volume law. At lower frequencies, the OPEE scales as $\mathcal{O}(L)$. The RMT value is shown as dashed lines. Other parameters: $J = 1$, $J_2 = 0.9$, $B_x = 0.1$.  }.\label{fig:OPEE}
\end{figure*}

\subsection{Operator entanglement entropy}

The OPEE quantifies the degree of operator non-locality by treating it as a state in a doubled Hilbert space.
Given an operator \( O \) acting on a Hilbert space of \( L \) spin-$\frac{1}{2} $ degrees of freedom (i.e., of dimension \( 2^L \)), we begin by \emph{vectorizing} it into a pure quantum state in a doubled Hilbert space \( \mathcal{H}_{\text{in}} \otimes \mathcal{H}_{\text{out}} \), each of dimension \( 2^L \). This is done by flattening \( O \) into a column vector using row-major order: $
|O\rangle = \frac{1}{||O||_\text{Fro}} \, \mathrm{vec}(O)$. 
This normalization ensures that \( \langle O | O \rangle = 1 \), allowing \( |O\rangle \) to be interpreted as a state in \( \mathcal{H}_{\text{in}} \otimes \mathcal{H}_{\text{out}} \cong \mathbb{C}^{2^L} \otimes \mathbb{C}^{2^L} \). We reshape the vector \( |O\rangle \) into a rank-\( 2L \) tensor with dimensions \( (2, 2, \dots, 2) \), corresponding to the physical spin degrees of freedom. We then bipartition the system into two subsystems, \( A \) and \( B \), and the tensor is transposed to group all indices in \( A \) and \( B \), and reshaped into a matrix \( M_{ab} \), where multi-indices \( a \) and \( b \) correspond to the configurations of subsystems \( A \) and \( B \), respectively. We finally perform a singular value decomposition (SVD) on the reshaped matrix: $M = U \Sigma V^\dagger$.
The singular values \( \{ \lambda_i \} \) in \( \Sigma \) correspond to the Schmidt coefficients of the bipartition \( A|B \). The reduced density matrix \( \rho_A \) has eigenvalues \( \lambda_i^2 \), and the von Neumann entanglement entropy~\cite{PhysRevX.8.021026} is computed as:
\begin{equation}
S[O] = - \sum_i \lambda_i^2 \log \lambda_i^2.
\end{equation}
Our numerical results for OPEE are shown in Fig.~\ref{fig:OPEE}. In panels (a) and (b), we show the evolution of the OPEE as a function of the RG time $\lambda$, for low $\Omega \lesssim \Omega_*$ and high frequency $\Omega\gtrsim \Omega_*$, where $\Omega_*$ is a system-size dependent ``threshold" scale. In panel (c), we show the asymptotic saturation value of the OPEE as $\lambda\to\infty$ as a function of $\Omega$.

As shown in Fig.~\ref{fig:OPEE}(a), when $\Omega\lesssim \Omega_*$ the OPEE appears to be insensitive to the difference between freezing and non-freezing points. Furthermore, the saturation value for OPEE agrees with that prescribed by random-matrix theory, with volume-law scaling~\cite{opee1}. On the other hand, for higher frequencies, the OPEE shows difference between freezing and non-freezing points as can be seen in Fig.~\ref{fig:OPEE}(b). In particular, the OPEE at the freezing point saturates to a lower value compared to that at the non-freezing point. In Fig~\ref{fig:OPEE}(c) we plot the saturation value of OPEE as a function of $\Omega$ across different system sizes. Moreover, this asymptotic saturation value deviates from the random-matrix prediction and exhibits a sub-volume-law scaling. This leads us to conjecture that the eventual thermalization behavior of the driven system, even at freezing, depends on how the drive frequency compares to a system-size dependent threshold, $\Omega_*(L)$. 

To further investigate the thermalization properties in the thermodynamic limit $L\to\infty$, we need to analyze the $L-$dependence of $\Omega_*$. Based on the previous observations, if $\Omega_*(L)$ diverges with $L$ in the thermodynamic limit, it suggests that an infinite system driven at a finite frequency always thermalizes. On the other hand, if $\Omega_*(L\to\infty)$ is finite, it suggests that even an infinite system can avoid thermalization if driven at a sufficiently high but finite frequency. While our current numerical results are limited in size by ED, finite-size scaling suggests that $\Omega_*$ increases slowly with $L$, and a more systematic future analysis is required to settle this question definitively. 

We note here that a quantum state at freezing can exhibit volume-law scaling of the {\it state} entanglement entropy \cite{AHaldar2021},  which has been attributed to Hilbert space fragmentation related to additional conserved quantities \cite{AHaldar2024}. The OPEE that is studied here is a distinct concept from the state entanglement, which quantifies the spread of operator in the Liouville space. Saturation to the full RMT value we observed here for $\Omega<\Omega_*(L)$ is a significantly stronger diagnostic than the observation of a volume law alone: it requires unrestricted mixing across the entire operator Hilbert space, implying that $H_0(\lambda)$ explores the full Liouville space without constraint.

Our study of the OPEE has also revealed two additional features tied to freezing (see \cite{supp} for details), which include: (1) The OPEE for higher-order freezing points display an additional oscillatory behavior for short RG time $\lambda$ in the ramp-up regime, and the number of oscillations exactly matches the order of the freezing point minus one, which is reminiscent of the ``dips" shown in Fig.~\ref{fig:freezing}(b). However, for later RG times the OPEE saturates to a value similar to the first freezing point, irrespective of the ratio of $\Omega/\Omega_*$. (2) The instanton events leave their fingerprints on the OPEE as ``step" like features as a function of $\lambda$.

\subsection{Diagonal ensemble average}

\begin{figure*}[t]
    \centering
    \includegraphics[width=1.0\linewidth]{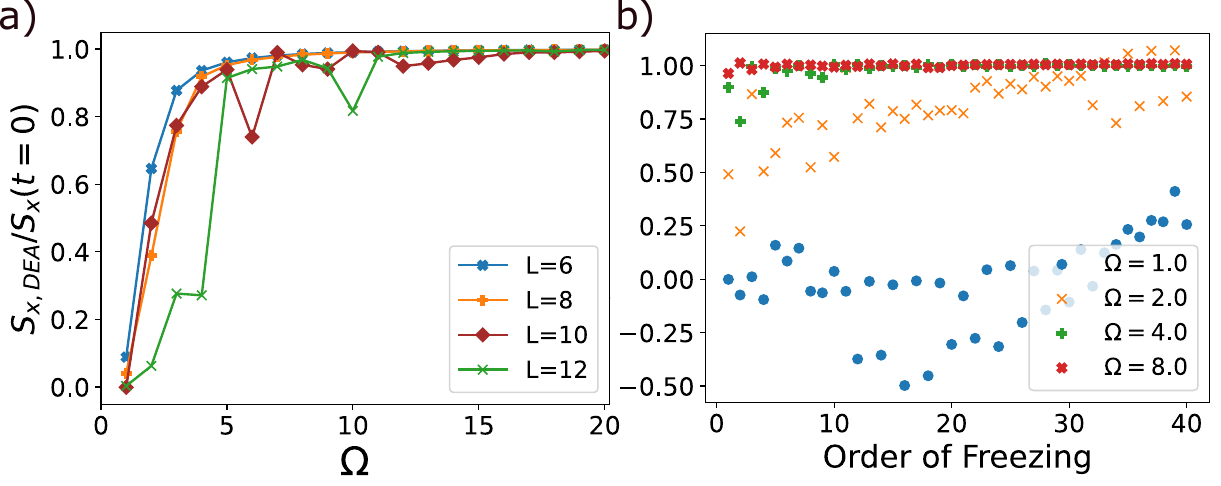}
    \caption{Diagonal ensemble average of $S_x$ for a thermal pure state as a function of (a) driving frequency for different system sizes, (b) order of freezing for different driving frequencies. Other parameters: $L=10$, $B_x=0.1$, $J_2=0.9$.} \label{deaverage}
\end{figure*}

To study how the memory of an initial state is preserved or lost for the driven Hamiltonian, we compute the diagonal ensemble average, which is intended to capture the expectation value of a local observable in the infinite-time limit. The DE average is defined as,
\begin{equation}
\begin{split}
\mathcal{O}_\text{DEA}&=\langle \Psi(NT\to\infty)| \mathcal{O}| \Psi(NT\to\infty) \rangle\\
&=\sum_{n} |c_{n}|^2 \langle \phi_{n}|\mathcal{O}| \phi_{n} \rangle    
\end{split}
\end{equation}
where $\ket{\phi_{n}}$ are the Floquet states and $\ket{\Psi(0)}=\sum_{n}c_{n} \ket{\phi_{n}}$ is an initial state written in the floquet basis. We will be interested in this quantity in the presence of a static magnetic field, $B_x$, which as was discussed previously improves the quality of freezing. We evaluate the DE average of $S_{x}$ starting from a thermal pure state.
We will first perform this calculation at the first-order freezing point while varying the driving frequency across different system sizes. Fig.~\ref{deaverage}(a) shows the normalized diagonal ensemble average $[S_{x,\mathrm{DEA}}/S_x(t=0)]$ as a function of $\Omega$ for several system sizes. At low drive frequencies ($\Omega/J \lesssim 2$), we find that this ratio is consistent with thermalizing behavior, with a value approaching $0$ and far below $1$, where the latter corresponds to perfect freezing (i.e., complete memory of the initial $S_x$ value). With increasing $\Omega$, the DE average at the first-order freezing point exhibit better memory of the initial state. Importantly, larger systems exhibit a delayed onset associated with this ratio of near-unity values, indicating that stronger driving is needed to stabilize freezing as the system size grows. For $\Omega/J \gtrsim 12$, all of the system sizes we have studied exhibit a value of this ratio close to $1$ ($>0.99$), consistent with near-perfect freezing. The small dips in the $L=10$ and $L=12$ data at intermediate $\Omega$ suggest finite-size resonances or residual heating processes that modestly reduce the quality of freezing. All of these trends align with the OPEE results of the previous subsection --- small $\Omega$ yields RMT-like operator entanglement and loss of memory of initial state at freezing, whereas large $\Omega$ suppresses saturation of OPEE and preserves initial state memory. The threshold frequency $\Omega_*$ between the two regimes weakly increases with system size $L$. In Fig.~\ref{deaverage}(b), we present results for the same ratio at a fixed, but few different values of $\Omega$ while increasing the freezing order. As expected based on the discussion up until now, we do not find the ratio to be close to $1$ for small $\Omega$ for even high freezing orders (up to the 40$^\text{th}$ order). Interestingly, for a fixed drive amplitude, larger frequencies yield higher DE average values than smaller ones. Therefore, even with the magnetic field present, freezing is not associated with an ``exact" conservation. This behavior stands in sharp contrast to results obtained with a square-wave drive \cite{AHaldar2024}, suggesting that these two driving protocols may belong to different universal classes.

\section{Analytical results}\label{sec:analytics}

In this section, we present analytical results related to the solutions of the flow equations Eq.~\eqref{eq:Flow}, especially in the putative dynamically frozen regime and beyond. Our discussion is divided into two parts. First, at early times, which includes the ramp-up regime and the approach to the prethermal fixed-point, the flow equation can be analyzed in a perturbative manner that is similar in spirit to the Floquet-Magnus expansion, but with important key differences that appear in the beyond leading-order terms. Second, in the late-time regime (which is the main regime of interest), the flow equation describes the departure from the prethermal fixed point and is analyzed non-perturbatively. This is where we obtain an approximate analytical handle on the instanton events that control the flow towards subsequent thermalization. 

\subsection{Early time: approach to prethermal behavior}

The early-time behavior of Eq.~\eqref{eq:Flow} with a fRG time, $\lambda\sim \Omega^{-1}$, is associated with the emergence of the prethermal behavior in the ramp-up regime, which is more pronounced in the high-frequency $\Omega\to\infty$ limit. Here, both $H_0(\lambda)$ and $H_1(\lambda)$ are expected to be exponentially local (operator strings of length $l$ will have a coefficient that decays exponentially in $l$). To describe the associated flow-dynamics, it is useful to perform a high-frequency expansion in terms of the initial Hamiltonian, $H_0(0)/\Omega$. Formally, we rescale the static component at $\lambda=0$ by $H_0(0)\to H_0(0)\zeta$, and then organize the Hamiltonian components for $\lambda>0$ in powers of $\zeta$:
\begin{subequations}\label{eq:Flowexpand}
    \begin{eqnarray}
        H_0(\lambda)&=& \zeta H_0^{(1)}(\lambda)+\zeta^2 H_0^{(2)}(\lambda)+\dots\,,\\
        H_1(\lambda) &=&  H_1^{(0)}(\lambda)+\zeta H_1^{(1)}(\lambda)+\dots\,.
    \end{eqnarray}
\end{subequations} 
Note that $\zeta$ simply serves as a convenient dimensionless (book-keeping) parameter; the actual expansion is organized in powers of $H_0(0)/\Omega$. 

The expansion in Eq.~\eqref{eq:Flowexpand} clearly shares some similarity with the standard Floquet-Magnus expansion, which is believed to capture the asymptotic high-frequency behavior of a Floquet system \cite{AEckardt2015}. For instance, the leading order term can be computed analytically, as discussed in detail in Appendix.~\ref{app:magnus}. Under the simplifying condition, $H_1(0)=H_{-1}(0)$, we obtain at leading order in $\zeta$:
\begin{subequations}\label{eq:Flow_pre}
    \begin{eqnarray}
        H_0^{(1)}(\lambda)&=&\left[J_1(\hat{z}e^{-\lambda \Omega})Y_0(\hat{z})-Y_1(\hat{z}e^{-\Omega\lambda})J_0(\hat{z})\right] \label{eq:H0lambda} \\
  &\times&\frac{\pi}{2}e^{- \Omega\lambda}\hat{z}H_0(0)\,, \nonumber\\
  H_1^{(0)}(\lambda)&=&H_1(0)e^{-\Omega\lambda}\,,\label{eq:H10lambda}
    \end{eqnarray}
\end{subequations}
where $\hat{z}=2\mathrm{Ad}_{H_1(0)}/\Omega$ is an operator that acts on $H_0(0)$, and $J_m$ and $Y_m$ are Bessel functions which are interpreted as a formal series in $\hat{z}$. We use the notation $\Ad_X$ as the adjoint operator: $\mathrm{Ad}_X Y=[X,Y]$.

The leading order solution in Eq.~\eqref{eq:Flow_pre} describes the approach to the prethermal fixed-point in the ramp-up regime. The typical fRG timescale associated with this approach is {\it rapid}, of ${\cal{O}}(\Omega^{-1})$. As we approach the fixed point, the oscillatory component $H_1(\lambda)$ decays exponentially with the fRG time $\lambda$, and the static component $H_0(\lambda)$ approaches a high-frequency effective Hamiltonian. 
The resulting effective Hamiltonian agrees with the (modified) Magnus expansion,
\begin{equation}\label{eq:H0mag}
  H_0^{(1)}(\lambda\to\infty)=H_F^{(0)}\equiv J_0\left(\frac{2\mathrm{Ad}_{H_1(0)}}{\Omega}\right)H_0(0)\,,
\end{equation} 
which serves as the leading order description of the prethermal fixed-point. Here, the $\lambda\to\infty$ limit can be refined to $\lambda\gg (1/\Omega)\ln(\Fnorm{H_1(0)}/(\Omega\Fnorm{\hat{1}}))$ based on Eq.\eqref{eq:Flow_pre}. Higher order corrections can be obtained order-by-order using the expansion in Eq.~\eqref{eq:Flowexpand}. However, we find that the higher order terms are {\it different} from the usual Magnus expansion. In particular, the correction term from the Magnus expansion for the present problem would be $\mathcal{O}(1/\Omega)$, but from the above flow expansion we obtain the correction to be $\mathcal{O}(1/\Omega^2)$.

The difference between these two approaches have a deeper explanation, that can be revealed from the point of view of a gauge dependence on the choice of $t_0$ --- the initial time for the drive cycle --- in Eq.~\eqref{eq:exactU}. In general, the Floquet Hamiltonian is gauge-dependent as a different choice of $t_0$  leads to different $H_F$, that is related via a gauge transformation. The Floquet-Magnus expansion directly approximates $H_F$ and therefore inherits the gauge dependence on $t_0$. The expansion yields $H_F=H_F^{(0)}+H_F^{(1)}+\mathcal{O}(1/\Omega^2)$. The leading order term $H_F^{(0)}$ is given by Eq.~\eqref{eq:H0mag}. The first appearance of an explicit gauge dependence in the Floquet-Magnus expansion occurs at the next-to-leading order, which takes the form
\begin{equation}\label{eq:HF1_main}
    H_F^{(1)}= -\sum_{m\neq 0 } \frac{1}{m\Omega}[h_m,h_0]\,.
\end{equation} 
Here $h_m$ is the Fourier coefficient of the co-moving frame Hamiltonian (see Appendix.~\ref{app:magnus} for more details). Note that Eq.~\eqref{eq:HF1_main} is explicitly gauge-dependent under a shift of $t_0\to t_0+\delta t$, where $h_m\to h_m \exp(im\Omega \delta t)$.  

On the other hand, the flow equation yields an approximation which is {\it manifestly gauge-invariant}, and is therefore able to capture many of the {\it intrinsic} gauge-invariant timescales in a more elegant fashion. This follows from the fact that the flow equation is invariant under $H_1\to e^{i\theta} H_1$, $H_0\to H_0$. Therefore, the static component $H_0(\lambda)$ represents a gauge-averaged approximation of $H_F$, and in hindsight, it is natural to expect a difference between the standard Floquet-Magnus expansion and the flow result.

 Let us now apply the solution in Eq.~\eqref{eq:Flow_pre} to explicitly address dynamical freezing. 
  The setup is as follows: we choose the bare drive Hamiltonian, $H_1(0)=H_{-1}(0)=A\calQ$. Here $\calQ$ is the charge of the emergent symmetry, and $A$ is the driving amplitude, as before. We choose the bare static Hamiltonian to be, $H_0(0)=H_{0,\calQ=0}+H_{0,\calQ=1}$. As the subscripts suggest, the two parts have zero and unit eigenvalues, respectively, under the action of $\calQ$: $[\calQ,H_{0,\calQ=q}]=q H_{0,\calQ=q}$. The static Hamiltonian associated with the prethermal fixed point is then,
 \begin{equation}
    H_0(\lambda_\text{pre})\approx H_{0,\calQ=0}+J_0\left(\frac{2A}{\Omega}\right) H_{0,\calQ=1}+\mathcal{O}\left(\frac{1}{\Omega^2}\right)\,. 
 \end{equation} Here, we have chosen to flow the system to $\lambda=\lambda_\text{pre}$, such that $\Omega\lambda_\text{pre}\gg 1$, but but still short compared to the thermalization timescales, which will be discussed later. 
 Therefore, by choosing $2A/\Omega=\nu_i$, where $\nu_i$ are zeros of the Bessel function $J_0(...)$, the prethermal Hamiltonian will display an (approximate) emergent symmetry $[H_0(\lambda_\text{pre}),\calQ]=\mathcal{O}(1/\Omega^2)$ in the high-frequency limit. To leading order, the locations of the freezing points agree with the prediction of the Magnus expansion.

 The leading order solution in Eq.~\eqref{eq:Flow_pre} also helps explains several features of the numerical results presented in Sec.~\ref{sec:spinchain}. First, in the approach to the prethermal Hamiltonian, the commutator $[H_0(\lambda),\calQ]$ shows dips whose number is given by $(n-1)$ for a freezing point, FP$_n$, respectively. Substituting $\hat{z}=\nu_i$ into Eq.~\eqref{eq:Flow_pre}, we see that the dips are associated with zeros of $J_1(...)$, which correspond to extreme points of $J_0$, and between every zero of $J_0$ there is one  extreme point. Second, to approach the prethermal fixed point, we require $(2A/\Omega)\exp(-\Omega\lambda)\ll 1$. Therefore, for higher freezing points corresponding to larger $A/\Omega$, it takes longer fRG time to reach the prethermal plateau.

However, as is well known, for a generic driven system that thermalizes to an infinite temperature state, the expansion in Eq.~\eqref{eq:Flowexpand} must eventually break down; the  higher-order terms in $H_0/\Omega$ should become increasingly relevant at a later fRG time. This connection between the breakdown of Eq.~\eqref{eq:Flowexpand} and heating can be proven based on our earlier discussion of Eq.~\eqref{eq:teff}. Suppose there is no breakdown, such that the exponential decay of $H_1(\lambda)$ implies that $\exp(-itH_0(\lambda))$ is an increasingly better approximation of the actual time-evolution operator. However, this contradicts heating because $\exp(-itH_0(\lambda))$ conserves energy. In other words, the prethermal fixed point is unstable in a generic non-integrable system. In the next subsection, we present a qualitative and semi-analytic treatment of the subsequent dynamics even at freezing, beyond the perturbative considerations of the present section.

\subsection{Late-time: thermalization via instantons}

To describe the nonlinear effects in Eq.~\eqref{eq:Flow}, consider running the flow equations up to some initial fRG time $\lambda_\text{pre}$, during which the prethermal flow in Eq.~\eqref{eq:Flow_pre} is a good approximation. Recall that $\lambda_\text{pre}$ satisfies $\lambda_\text{pre}\gg \Omega^{-1}$, but is smaller than the {\it a priori} unknown thermalization fRG time. We also assume that heating is sufficiently slow so that  $\lambda_\text{pre}$ satisfies $\Fnorm{H_1(\lambda_\text{pre})}\ll \Fnorm{H_0(\lambda_\text{pre})}$. Starting from $\lambda=\lambda_\text{pre}$, we can integrate Eq.~\eqref{eq:FlowH1} by assuming $H_0(\lambda)\approx H_0(\lambda_\text{pre})$, yielding
\begin{equation}\label{eq:H1_linearized}
\begin{split}
    &H_1(\lambda)\approx e^{-\Omega(\lambda-\lambda_\text{pre})}\times \\
    &e^{-(\lambda-\lambda_\text{pre})H_0(\lambda_\text{pre})}H_1(\lambda_\text{pre})e^{(\lambda-\lambda_\text{pre})H_0(\lambda_\text{pre})}\,.
\end{split}
\end{equation} 
Expanding Eq.~\eqref{eq:H1_linearized} in the eigenbasis $(E_m,\ket{m})$ of $H_0(\lambda_\text{pre})$, we find
\begin{equation}\label{eq:H1_matelem}
    \braket{m|H_1(\lambda)|n}\approx e^{(-\Omega-E_m+E_n)(\lambda-\lambda_\text{pre})} \braket{m|H_1(\lambda_\text{pre})|n}\,.
\end{equation}

According to Eq.~\eqref{eq:H1_matelem}, $H_1(\lambda)$ will decay if $\Omega>|E_m-E_n|$ for all pairs of $\{m,n\}$ with $\braket{m|H_1(\lambda_\text{pre})|n}\neq 0$. This case corresponds to the pre-thermal fixed point being stable, and the associated expansion in Eq.~\eqref{eq:Flowexpand} is expected to converge. However, this condition is not always satisfied, even for exponentially local Hamiltonians. Taking a spin chain as an example, an exponentially local drive term $H_1(\lambda_\text{pre})$ can contain terms that flip $l$ spins simultaneously for arbitrary $l$, despite its amplitude decaying exponentially in $l$. If the typical energy of a single spin flip is $J$, the matrix elements for $l>\mathcal{O}(\Omega/J)$ spin flips will eventually grow after a sufficiently long fRG time. In Appendix.~\ref{app:timescale}, we provide an estimate for these time scales associated with the evolution of $\Fnorm{H_1(\lambda)}$. 

We argue that $\Fnorm{H_1(\lambda)}$ is expected to show a minimum at $\lambda=\lambda_\text{min}\sim (1/J)\ln(\Omega/J)$, where $\Fnorm{H_1(\lambda)}\sim \exp\left[-(\Omega/J)\ln(\Omega/J)\right]$; see Fig~\ref{fig:L10thermalization} in Appendix \ref{app:timescale} for numerical evidence of such scaling. This implies that we can choose $\lambda_\text{pre}=\lambda_\text{min}$, and $H_0(\lambda_\text{min})$ is a good approximation of the time evolution up to exponentially long time in $\Omega$. This is consistent with generic expectations that heating in rapidly driven system is exponentially slow in $\Omega$ \cite{prethermal1,prthermal2,weidinger2017floquet,prethermal3,PhysRevLett.120.197601,prethermal4,prethermal6,prthermal5,PhysRevLett.132.100401} . When $\lambda\gg \lambda_\text{min}$, we expect a growth in $\Fnorm{H_1(\lambda)}$, where different operators in $H_1(\lambda)$ are expected to have distinct thermalization timescales. For an operator that flips $l$ spins, the thermalization time scale is expected to be 
\begin{equation}
    \lambda^{(l)}_\text{th}\sim \frac{f(l)+l\ln(\Omega/J)}{lJ-\Omega}\,,
\end{equation} 
where the function $f(l)$ is a dimensionless function of $l$ that depends on the microscopic details of the system. The time-scale $\lambda_\text{th}^{(l)}$ is such that at $\lambda=\lambda_\text{th}^{(l)}$, the matrix element of $H_1(\lambda)$ corresponding to $l$ spin flips becomes $\mathcal{O}(1)$.  As discussed before, only operators that have grown to be longer than $l>\mathcal{O}(\Omega/J)$ lead to that associated thermalization.  For $l\gg \mathcal{O}(\Omega/J)$, $\lambda_\text{th}^{(l)}$ is parametrically similar to $\lambda_\text{min}$.

We can describe the evolution of $H_1(\lambda)$ approximately, under the assumption that only one of the matrix elements $\braket{m|H_1(\lambda)|n}$ grows to be $\mathcal{O}(1)$ --- we dub this a single-instanton event. In this scenario, we ignore all of the other matrix elements of $H_1(\lambda)$. Then, the term $[H_1,H_1^\dagger]$ in Eq.~\eqref{eq:FlowH0} becomes approximately diagonal in the instantaneous eigenbasis of $H_0$. The flow equations simplify to
\begin{subequations}\label{eq:flow_nonlinear}
\begin{eqnarray}
    \partial_\lambda E_m(\lambda) &=& 2 |\braket{m|H_1(\lambda)|n}|^2\,, \\
    \partial_\lambda E_n(\lambda) &=& -2 |\braket{m|H_1(\lambda)|n}|^2\,, \\
    \partial_\lambda \braket{m|H_1(\lambda)|n} &=& [-\Omega-E_m(\lambda)+E_n(\lambda)]\braket{m|H_1(\lambda)|n} \nn\,. \\
\end{eqnarray}
\end{subequations} 
Here $E_m(\lambda)$ and $E_n(\lambda)$ are the eigenvalues of the instanteous eigenvectors $\{\ket{m},\ket{n}\}$ of $H_0(\lambda)$, and $\ket{m},\ket{n}$ are approximately independent of $\lambda$. The analytical solutions are given by,
\begin{subequations}\label{eq:nonlinear_solution}
\begin{eqnarray}
    E_n(\lambda)-E_m(\lambda) &=& \Omega - \tilde{\Omega} \tanh(\tilde{\Omega}(\lambda-\tilde{\lambda}))\,, \\
    |\braket{m|H_1(\lambda)|n}| &=& \frac{\tilde{\Omega}}{2\cosh (\tilde{\Omega}(\lambda-\tilde{\lambda}))}\,, \label{eq:cosh}
\end{eqnarray}
\end{subequations} where $\tilde{\Omega},~\tilde{\lambda}$ are integration constants determined from initial conditions at $\lambda=\lambda_\text{pre}$. Several remarks are now in order.

For a generic non-integrable Hamiltonian, we expect the above single-instanton scenario to be a good approximation for the thermalization dynamics, in the absence of any special degeneracies in $E_m(\lambda)-E_n(\lambda)$, and when only one pair of levels are coupled strongly by $H_1(\lambda)$ via $\braket{m|H_1(\lambda)|n}$. Under this assumption, different matrix elements of $H_1(\lambda)$ will become ${\cal{O}}(1)$ at different fRG times, and at a given fRG time only one particular matrix element is large. As a corollary, near $\lambda=\tilde{\lambda}$, the norm $\Fnorm{H_1(\lambda)}$ will be dominated by a single matrix element. This line of reasoning and the associated analytical form of $\Fnorm{H_1(\lambda)}$ in Eq.~\eqref{eq:cosh} is consistent with our numerical results, as shown by the excellent fit in Fig.~\ref{fig:H1nL10} (Additional examples are also shown in Appendix~\ref{app:Supp}.). We note that in the presence of additional symmetries that enforce degeneracies in $E_m(\lambda)-E_n(\lambda)$, the Eqs.~\eqref{eq:flow_nonlinear} can be extended to include the whole irreducible representation of the symmetry. We leave this generalization for future work.

Due to the explicitly broken time-translation symmetry, the Floquet Hamiltonian is only well defined modulo $\Omega$. However, at $\lambda=0$ the initial Hamiltonian $H_0(0)$ can have a many-body bandwidth larger then $\Omega$. The instanton solution in Eq.~\eqref{eq:nonlinear_solution} connects these two limits by describing the band-folding dynamics of $H_0(\lambda)$; see e.g. the schematic in Fig.~\ref{fig:Schematic}. During the instanton event, $(E_n-E_m)$ decreased from $\Omega+\tilde{\Omega}$ to $\Omega-\tilde{\Omega}$. If $|\tilde{\Omega}|>\Omega$, there will appear to be an accidental degeneracy at $\lambda=\tilde{\lambda}$. We expect that this degeneracy will be lifted after including the other matrix elements neglected in Eq.\eqref{eq:flow_nonlinear}. Additionally, we note that when $|\Omega-\tilde{\Omega}|>\Omega$, a single instanton event is not enough to fold $(E_n,E_m)$ into an energy interval shorther than $\Omega$. In that case, this can trigger subsequent instanton events driven by the matrix element $\braket{n|H_1(\lambda)|m}$ (note that $m,n$ is exchanged), where now it satisfies the growth criterion $E_m-E_n>\Omega$.
 On the contrary, there will be no subsequent instanton events in $\braket{n|H_1(\lambda)|m}$ nor $\braket{m|H_1(\lambda)|n}$ if $|\Omega-\tilde{\Omega}|<\Omega$. Therefore, the energy levels of $H_0(\lambda)$ are folded step by step in a series of instanton events. When no further instanton events occur, $H_0(\lambda)$ has the property that $|E_m-E_n|<\Omega$ for any pair of eigenstates $m,n$, and at the same time $H_1(\lambda)$ will decay exponentially to zero.

\section{Outlook}
\label{sec:outlook}
Going beyond conventional Floquet-Magnus high-frequency expansions and numerical exact diagonalization of microscopic Hamiltonians, here we have employed the Floquet flow formalism to analyze the phenomenology of dynamical freezing from a fresh (and universal) perspective. Most interestingly, we find that the bare driven Hamiltonian flows through a complex landscape of (unstable) fixed points and shows a strong tendency towards {\it slow} thermalization at the dynamically frozen points. Freezing itself is associated with an emergent approximate conservation law, that appears at the first prethermal fixed point along the flow. By combining numerical solutions of the flow equations using exact diagonalization and matrix-product operator-based techniques, as well as analytical approximations that are non-perturbative in the inverse drive frequency ($\Omega$), we have shown that freezing is approximate at any finite $\Omega$. Moreover, the violation of the emergent conservation law, quantified by the commutator $[H_0(\lambda),\mathcal{Q}]$, vanishes as $1/\Omega^2$ in the high-frequency limit. 

One of the key contributions of this work is the analysis of thermalization dynamics of the system beyond the prethermal fixed point, where we find that the flow between the intermediate fixed points can be described using {\it instanton} events. For the sake of concreteness, we have chosen to study the driven spin chain as our main numerical example using complementary methods. However, the demonstrated features for freezing and thermalization are quite generic, as demonstrated by the beautiful agreement between our numerical results and the general model-independent analytical computations. Using a combination of different numerical diagnostics, we have demonstrated that the system flows to a random-matrix-like Hamiltonian when the frequency $\Omega$ is below a system-size-dependent threshold, $\Omega_*(L)$. Due to present numerical limitations of simulating only small system sizes, we have only been able to obtain the weakly increasing $L-$dependence of $\Omega_*(L)$ in a narrow range. Further numerical and analytical advances will be required to obtain the form of  $\Omega_*(L)$ for a generic non-integrable Hamiltonian. If $\Omega_*(L)$ diverges with $L$ in the thermodynamic limit, that indicates fully thermalizing behavior even at dynamical freezing. However, it is possible that dynamical freezing may slow down the rate of divergence, or keep $\Omega_*(L)$ to be finite in the $L\rightarrow\infty$. This remains an interesting open problem for the future.

From our perspective, the fundamental new ingredient offered by the flow formalism at freezing analysed here correspond to the instanton events. We speculate that the instantons correspond to the real-time dynamics in the following sense: In the high-frequency limit, the system starts to exhibit heating after a time, $t_\text{th}\sim (\Omega/J)^{\Omega/J}$. However, the energy absorption is not a continuous process, and the system can only absorb a quantum of $\Omega$ each time.  The energy absorption process is associated with a transition between two eigenstates $\ket{m},\ket{n}$ of the prethermal Hamiltonian whose energy differs by integer multiples of $\Omega$ \cite{PhysRevLett.132.100401}, and the transition is alternatively described as the hybridization between $\ket{m}$ and $\ket{n}$ due to the drive. We speculate that (a) the instanton event effectively implements the hybridization process, by mixing the eigenvectors in a step-by-step fashion. In addition, the energy absorption events are expected to be rare in real time, and the typical interval between them is of order $t_\text{th}$. This implies that after absorbing one energy quantum, the system should enter a new ``prethermal" regime as the next absorption event is $t_\text{th}$ away. To that end, we conjecture that (b) the intermediate fixed points in between the instantons are the effective Hamiltonians for these new ``prethermal" regimes. To verify this picture, it requires simulations of real-time dynamics up to exponentially long times, which is a challenging, but exciting future direction.

We note that despite the simple (yet elegant) solution in Eq.~\eqref{eq:nonlinear_solution} for a single instanton event, the thermalization dynamics is still quite complex and ``chaotic" in nature. There are several aspects that our simple analytical treatment does not capture: First, in the single-instanton scenario, we have only focused on a single large matrix element which grows to $\mathcal{O}(1)$ but ignored the others, which is sufficient to describe the current instanton event. However, the other possibly small matrix elements are likely strongly perturbed by the presence of the instantons, which substantially affects the following instanton events. Second, while we have argued that single-instanton events are more likely, it is possible that two instanton events happen in relatively quick succession (e.g. in Fig.~\ref{fig:H1nL10} (a)). We leave a detailed analytical study of such two- and multiple-instanton events for future investigation.  

We conclude by discussing several additional future directions. 
In the current work, we have only considered the cosine wave drive for simplicity, as there are just two Fourier components in the flow equation. Recently, there has been numerical evidence suggesting that dynamical freezing is more robust for square-wave drive \cite{AHaldar2021,AHaldar2024}. The flow formalism can be generalized to describe the square-wave drive, by introducing an infinite series of Hamiltonian components corresponding to the Fourier decomposition of the drive function \cite{MClaassen2021}. Whether the flow away from the first prethermal fixed point is similar in character to the example studied here, but with possibly delayed timescales, is an interesting open question.

Time evolution of matrix-product states and operators is in general computationally demanding, due to operator spreading and entanglement growth. In our setup, this problem is mitigated by dynamical freezing, where the slower entanglement growth compared to non-freezing points enabled simulation upto timescales long enough to access the prethermal regime. In the future, it would be useful to combine constraints imposed by the emergent symmetry at freezing with the recently developed MPO compression algorithm \cite{DEParker2020} to enable numerical simulation upto even longer timescales. Studying the operator spreading using these algorithms, both at and away from freezing, remains an interesting future direction.

\acknowledgments 

DC thanks Vedika Khemani for useful discussions, and Dan Mao for related collaborations. We thank Zihao Qi for discussions regarding the isospectral nature of the many-body spectrum during the flow. RM, HG and DC are supported in part by a New Frontier Grant awarded by the College of Arts and Sciences at Cornell University, and by a Sloan research fellowship from the Alfred P. Sloan foundation to DC. Some of the computations presented in this work are also supported in part by a NSF CAREER grant (DMR-2237522) to DC. HG is also supported by a Bethe-Wilkins-KIC postdoctoral fellowship at Cornell University.

\section*{DATA AVAILABILITY}
All the theoretical data generated in this study have been deposited at Zenodo and are publicly available as of the date of publication: \url{https://zenodo.org/records/18421731}.

\appendix 

\section{Connection between flow-renormalization and Schrieffer-Wolff transformation}\label{Supp:SW}

In this section, we provide a brief pedagogical review of the connection between the flow-renormalization approach \cite{kehrein2007flow}, and the more standard Schrieffer-Wolff (SW) transformation applied to time-independent Hamiltonians \cite{SW}. The SW transformation is often used to perturbatively eliminate the ``off-diagonal" high-energy terms in a Hamiltonian, resulting in an effective Hamiltonian within a low-energy subspace. The flow-RG equations essentially use a sequence of such infinitesimal SW transformations to gradually achieve this decoupling over a finite energy interval at a time. The SW transformation defines a unitary transformation \( U = e^S \), where \( S \) is an anti-Hermitian operator (i.e., \( S = -S^\dagger \)). The transformed Hamiltonian \( H' \) is given by:
\beq
H' = e^S H e^{-S} = H + [S, H] + \frac{1}{2}[S, [S, H]] + \dots,
\eeq
where \( S \) is chosen to eliminate the terms that couple different subspaces (e.g., low-energy and high-energy subspaces).

Typically, we write the original Hamiltonian as \( H = H_0 + V \), where \( H_0 \) is the unperturbed part and \( V \) is a perturbative coupling term. Then, \( S \) is chosen such that,
\beq
[S, H_0] = -V
\eeq
This condition ensures that the transformed Hamiltonian \( H' \) is approximately block-diagonal to a certain order in \( V \), decoupling the low-energy and high-energy subspaces.

Within the flow equation approach \cite{WG1,WG2,WG3,kehrein2007flow,PhysRevB.106.115440}, the Hamiltonian \( H \) is transformed continuously by defining a family of Hamiltonians \( H(l) \), where \( H(0) = H \) and \( H(l) \) flows toward a diagonal (or block-diagonal) form as \( l \to \infty \). The flow is governed by the differential equation,
\beq\label{eq:A3}
\frac{dH(l)}{dl} = [\eta(l), H(l)],
\eeq
where \( \eta(l) \) is the generator of the transformation, chosen to be
\beq
\eta(l) = [H_{\text{diag}}(l), H_{\text{off-diag}}(l)].
\eeq
Here, \( H_{\text{diag}}(l) \) and \( H_{\text{off-diag}}(l) \) are the diagonal and off-diagonal parts of \( H(l) \), respectively. This choice of \( \eta(l) \) ensures that the flow gradually drives \( H(l) \) towards a diagonal form as \( l \to \infty \), effectively decoupling the low-energy and high-energy sectors.

The ideas discussed above can be generalized to the Floquet context using the Sambe representation \cite{SambeOriginal,MClaassen2021}. In this representation, the time-dependent Hamiltonian ${H(t)}=\sum_m e^{im\Omega t}H_m$ is represented in an extended Hilbert space, where the Floquet operator $H(t)-i\partial_t$ is represented as
\begin{equation}
\calH = \sum_m \left[m\Omega \otimes \ket{m}\bra{m}+\sum_M H_M\otimes \ket{m+M}\bra{m}\right]\,.
\end{equation} The Sambe representation has a gauge-redundancy, so that the eigenvalues of $\calH$ can be labeled as  
\begin{equation}
    \calH \ket{\phi_{nM}}=(\varepsilon_n+M\Omega)\ket{\phi_{nM}}\,,
\end{equation} for any $M\in \mathbb{Z}$. In particular, $\varepsilon_n$ are the eigenvalues of the Floquet Hamiltonian, $H_F$. 

Following Ref.~\cite{MClaassen2021}, the Floquet flow is constructed as a unitary flow in the Sambe representation, which automatically preserves the eigenspectrum of $\calH$ and hence the spectrum of $H_F$. To ensure unitarity, the flow should be generated by a commutator between the generator $\eta(l)$ and $\calH(l)$ as in Eq.~\eqref{eq:A3}. Additionally, the generator should be chosen such that it strives to remove the oscillatory parts of $\calH$ and it does not generate higher harmonics not present in $\calH$. The generator that satisfies the above requirements is
\begin{equation}
    \eta(l) = \sum_{m}\sum_{M>0} \left[H_M(l)\otimes\ket{m+M}\bra{m}-h.c.\right]\,.
\end{equation} 
When there are only two Fourier-components $H_0$ and $H_1$, the flow reduces to Eq.\eqref{eq:Flow} in the main text. 

\section{Additional numerical details}\label{app:numericaldetails}

Here we provide some additional details about our numerical simulations. In our simulations, we did not explicitly fix a maximum bond dimension; instead, we used the default singular value cutoff in ITensors.jl, which is $10^{-12}$. For all numerical results involving MPO evolution, the bond dimensions of both $H_{0}(\lambda)$ and $H_{1}(\lambda)$ remained below 500. To compute the entanglement entropy, we initialized the MPS in a fully polarized product state along the $x$-direction and evolved it using the TDVP method. Since this initial state is unentangled, we employed the global-Krylov method to enrich the MPS with Krylov vectors. For the time evolution of the MPS, we employ the default single-site TDVP algorithm; for further details, see~\cite{tensornetwork}. During the time evolution of the MPS, the SVD cutoff was set to $10^{-10}$. For the time evolution using the RK4 method, we set the integration step size to $10^{-2}$ for both the exact and MPS-based calculations. When studying the $\Omega$-scaling of the commutator at higher frequencies—where the associated time period becomes shorter—we reduce the step size to $10^{-3}$ to accurately resolve the faster dynamics.

\section{Additional numerical results}\label{app:Supp}

In this appendix, we present additional numerical results that support many of the important conclusions presented in the main text.
\begin{itemize}[leftmargin=*]
    \item {\bf Effect of static transverse magnetic field:}

In continuation of the discussion in Sec.~\ref{sec:transverse}, in Fig.~\ref{magnetization:App}(a), we plot the expectation value of $S_x$ in the eigenbasis of $H_0(\lambda_c)$, both with and without $B_x$, at and away from the freezing point. At the freezing point, the effective Floquet Hamiltonian to leading order is proportional to $YY + ZZ$ terms; see Eq.~\eqref{Eq:effective} in Appendix \ref{app:Magnus_spinchain}. This structure gives rise to an emergent $\mathbb{Z}_2$ symmetry, resulting in degenerate Floquet eigenstates (equivalently, eigenstates of $H_0(\lambda_c)$), and consequently, the expectation values of $S_x$ are nearly zero for all states. Introducing a small nonzero $B_x$ lifts the degeneracy, making the Floquet eigenstates to align with the eigenstates of $S_x$. In Fig.~\ref{magnetization:App}(b), we show the real-time magnetization profile for a thermal pure state evolving under $H_0(\lambda_c)$. 

We find that the presence of $B_x$ enhances the quality of freezing. In Fig.~\ref{fig:MagL10latetime} we show late (real-)time magnetization at the first freezing point for different frequency. Even though at the freezing point the dynamics ``slows down" the magnetization still decays, indicating only an approximate conservation law. 
In this simulation, $\Omega$ is chosen large enough to avoid resonances between the eigenstates of the unperturbed part of the Hamiltonian. For smaller driving frequencies (but $\Omega\gtrsim J$), even at the first freezing point, the magnetization decays substantially. For the same driving frequency, at a higher-order freezing point (blue curve, $A=15.32$) compared with the lower-order case (yellow curve, $A=6.01$), the magnetization still decays from its initial value. This indicates that while the presence of a magnetic field enhances freezing at the level of the expectation value of $S_x$, it cannot prevent the eventual decay; see also the section on diagonal ensemble average in the Supplementary information for a related discussion. Whether this decay ultimately leads to thermalization to infinite temperature in the $L\rightarrow\infty$ limit remains an open question.

\begin{figure}[htb]
    \centering
    \includegraphics[width=1\linewidth]{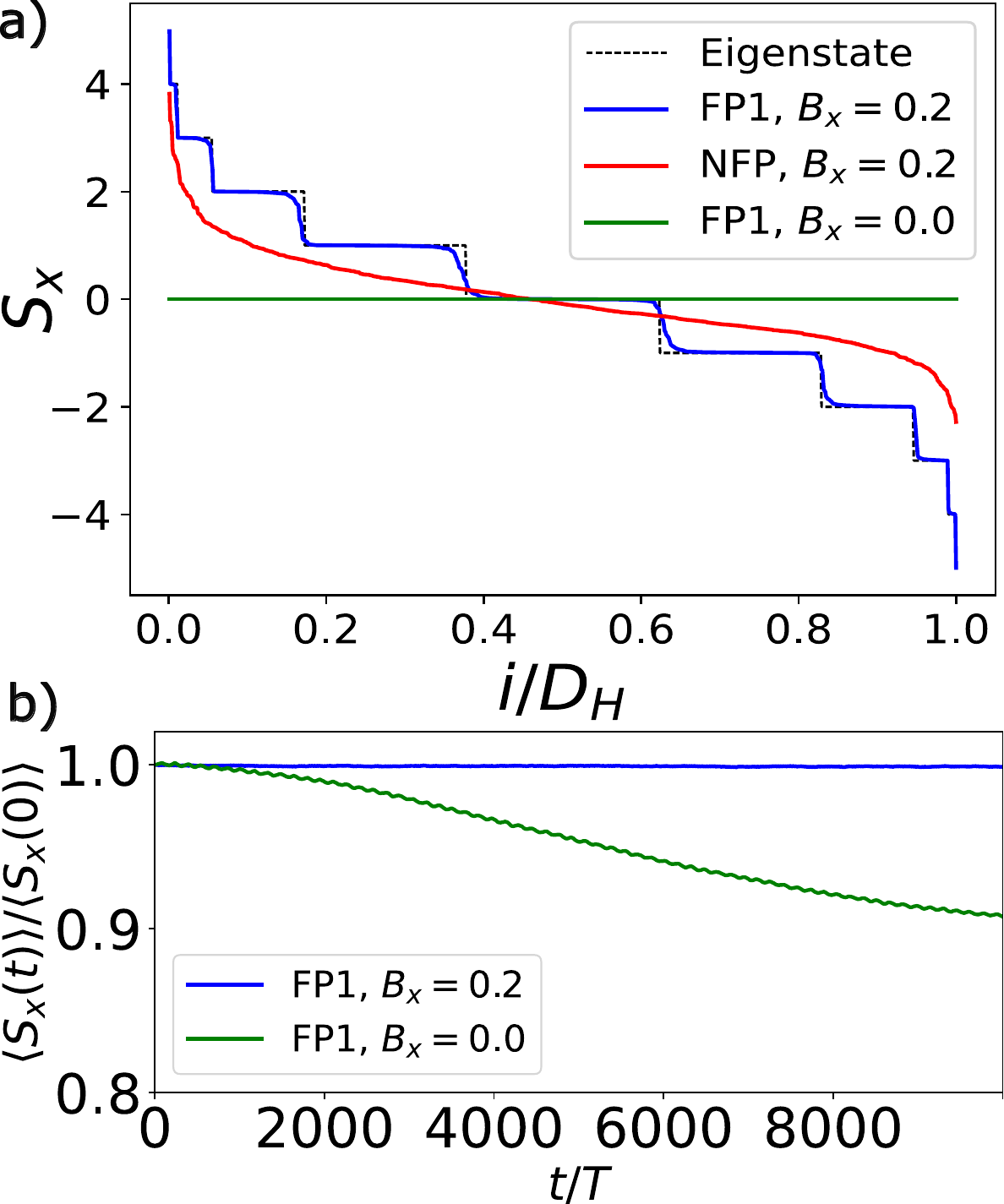}
    \caption{(a) The values of $S_x$ for the eigenstates of $H_{0}(\lambda_c)$ are plotted in decreasing order, comparing the freezing point (FP1) and the nonfreezing point (NFP), both with and without the magnetic field term in the $x$ direction. Without the magnetic field in the $x$ direction, at the freezing point, the leading-order effective Floquet Hamiltonian contains terms proportional to $YY+ZZ$, which implies a global $Z_2$ symmetry under $X_i \to -X_i$. As a result, the expectation value of $S_x$ in the eigenstates of $H_{0}(\lambda_c)$ is vanishingly small. Parameters: $L=10$, $J=1$, $J_2=0.2$, $\Omega=5$, $\lambda_c=5.0$. (b) Magnetization is plotted as function of real time at the first freezing point with/without the magnetic field term $B_x$. Parameters: $L=10$, $J=1$, $J_2=0.5$, $\lambda_c=5.0$.}\label{magnetization:App}
\end{figure}   

\begin{figure}[htb]
    \centering
    \includegraphics[width=1.0\linewidth]{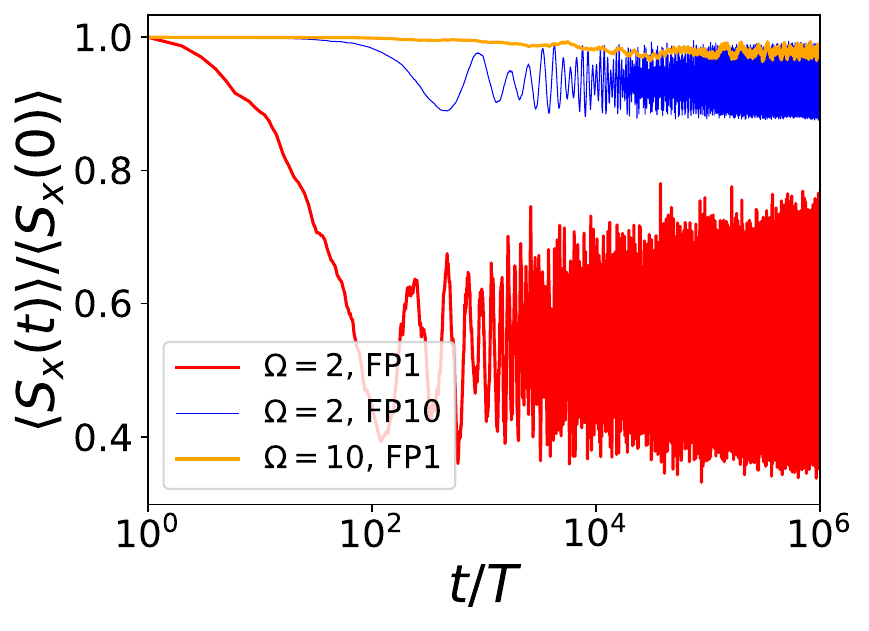}
    \caption{Magnetization for a thermal pure state (evolving under Hamiltonian $H_{0}(\lambda_c)$) as a function of real time at the first freezing point for two different driving frequencies. Parameters: $L=10$, $J=1$, $J_2=0.5$, $B_x=0.1$, $\lambda_c=10$.}
    \label{fig:MagL10latetime}
\end{figure}

\begin{figure}[htb]
    \centering
    \includegraphics[width=1.0\linewidth]{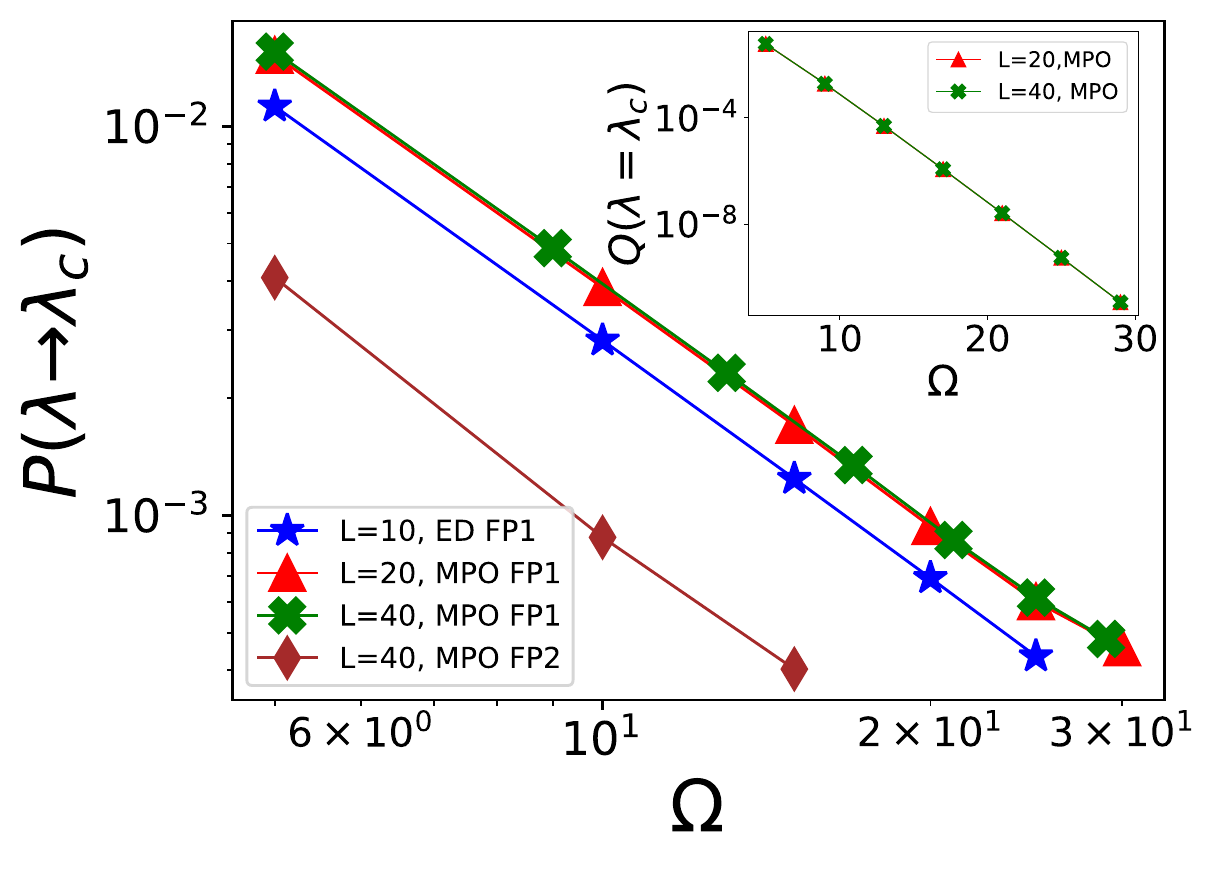}
    \caption{$P(\lambda \to \lambda_{c})$ for (\( L=10,~20,~40 \)) at the first freezing point (FP$_1$) exhibits a universal \( 1 / \Omega^2 \) scaling. Inset: \( Q(\lambda=\lambda_c) \) for different system sizes exhibits an exponential decay with increasing driving frequency. The data near FP$_2$ also follows the same frequency scaling.}\label{fig:frequencyscaling}
\end{figure}

\item{\bf Frequency scaling of $P(\lambda_c)$:}

Recall the definition of $P(\lambda)$ in Eq.~\eqref{commH0Xnorm}, which helps identify the extent of freezing and the emergent conservation law.  In Fig.~\ref{fig:frequencyscaling}, we plot $P(\lambda_c)$ at two distinct freezing points with increasing $\Omega$, where $\lambda_c$ is associated with the initial prethermal plateau. Interestingly, we find that $P(\lambda_c)$ decays as $1/\Omega^2$ at large $\Omega$, which is consistent with our analytical results in Sec.~\ref{sec:analytics}. Relatedly,  $Q(\lambda)$ (Eq.~\eqref{H1norm}) falls off exponentially with the drive frequency, $\Omega$ (Fig.~\ref{fig:frequencyscaling} inset). We also point out that $P(\lambda)$, which is normalized to be an intensive quantity, already appears to show convergence to the thermodynamic limit for $L=20$, given the lack of any appreciable difference from the $L=40$ results.

\item {\bf Dynamical freezing over longer fRG times:}

In the main text, we showed results for the plateau associated with $P(\lambda)$ upto $\lambda\sim1$. However, the plateau extends until later fRG times, as shown in Fig.~\ref{L10latetimecomm}, both at and away from freezing point for $\Omega=10$.

\begin{figure}[htb]
    \centering
    \includegraphics[width=1\linewidth]{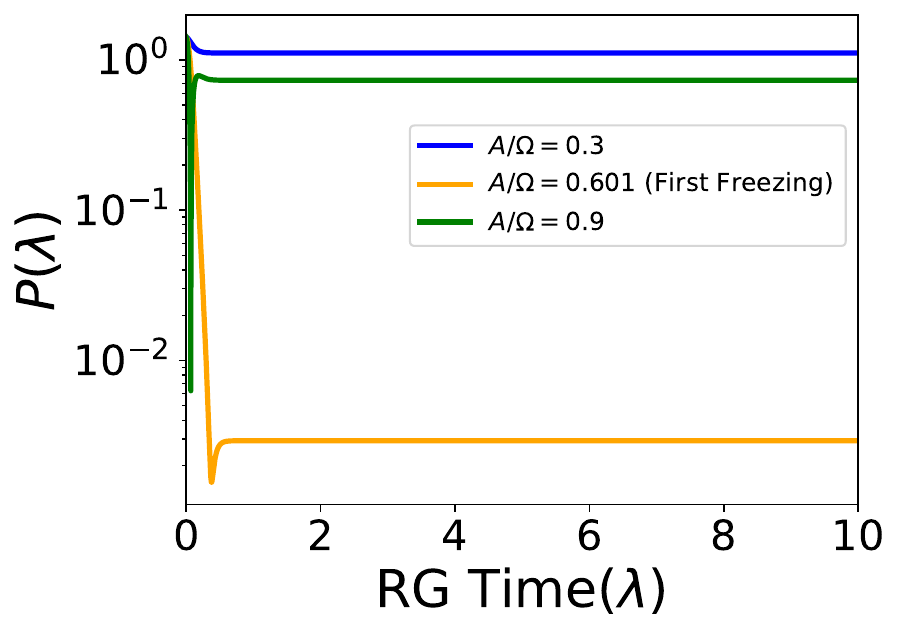}
    \caption{Normalized commutator, $P(\lambda)$, for $L=10$ as a function of $\lambda$ at and away from freezing. All other parameters are same as in Fig~\ref{fig:freezing}.}\label{L10latetimecomm}
\end{figure}

\item {\bf Renormalization of freezing points:} 

To examine the possible renormalization of the freezing point away from the leading-order Floquet-Magnus predictions, we compute numerically the location of the minima of \( P(\lambda_{c}) \) for various driving frequencies. In Fig.~\ref{fig:Supprenoormalization}, a fine-grid scan of \( P(\lambda_{c}) \) is performed near FP$_1$ as a function of \( A/\Omega \) for two different system sizes (\( L=8,~10 \)) using flow ED. We observe that with increasing frequency, the minimum of \( P(\lambda_{c}) \) approaches the value (\( =0.601 \)) predicted by the zeroth-order Magnus expansion. However, for lower frequencies, there is a slight shift in the location of the freezing point.

\begin{figure}[htb]
    \centering
    \includegraphics[width=1\linewidth]{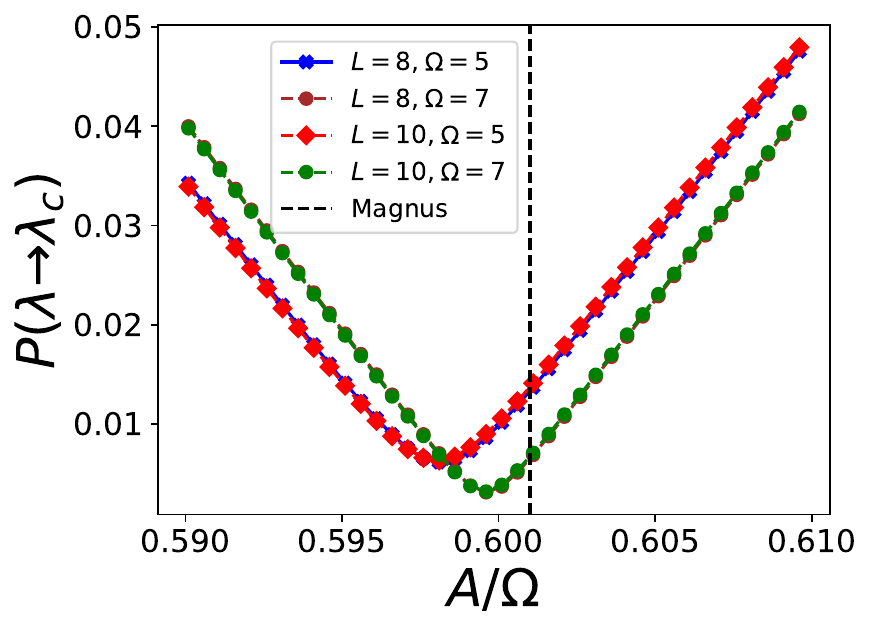}
    \caption{A fine-grid scan for \( P(\lambda \to \lambda_{c}) \) around FP$_1$, for (\( L=8,~10 \)) and two different driving frequencies. The dashed line represents the prediction from the zeroth-order Magnus term.}\label{fig:Supprenoormalization}
\end{figure}

\item {\bf Higher order freezing points:} 

In Fig.~\ref{fig:higherorder}, we illustrate the scenario where the driving frequency is comparable to the single-particle energy scales (in contrast to the choice of $\Omega$ in Fig.~\ref{fig:freezing}a), and the driving amplitude represents the largest energy scale in the system—effectively exploring higher-order freezing points while keeping the frequency fixed. In this case, we did not observe any qualitative differences; as the amplitude increases, the plateau value of $P(\lambda)$ decreases.

\begin{figure}
    \centering
    \includegraphics[width=0.9\linewidth]{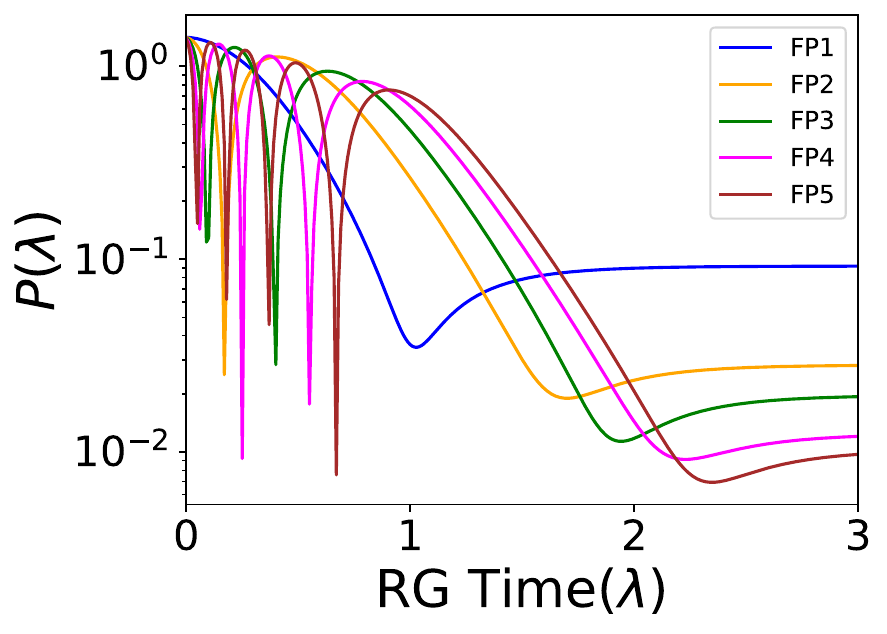}
    \caption{$P(\lambda)$ is plotted as a function of RG time for different freezing points, where the value of the asymptotic value decreases continuously for higher order freezing points. 
    Parameters: $L$=10, $J$=1, $J_{2}=0.2$, $\Omega=2.0$. }
    \label{fig:higherorder}
\end{figure}

\item {\bf Multiple separated instanton events:} 

In Fig.~\ref{fig:L6rf}(a), we plot the norm of \( H_{1} \) as a function of fRG time using flow ED for a system of size $L=6$. The two well-separated instanton events are clearly visible, where the norms of \( H_{1}(\lambda) \) peak at approximately \( \lambda \approx 125 \) and \( \lambda \approx 380 \), respectively. In Fig.~\ref{fig:L6rf}(b), the corresponding norm of \( H_{0}(\lambda) \) is shown, displaying a clear step-like feature precisely at the locations where \( H_{1}(\lambda) \) peaks. Calculations for longer fRG times are constrained by our numerical precision, which in this case is double precision (binary64).

\begin{figure}[htb]
    \centering
    \includegraphics[width=1\linewidth]{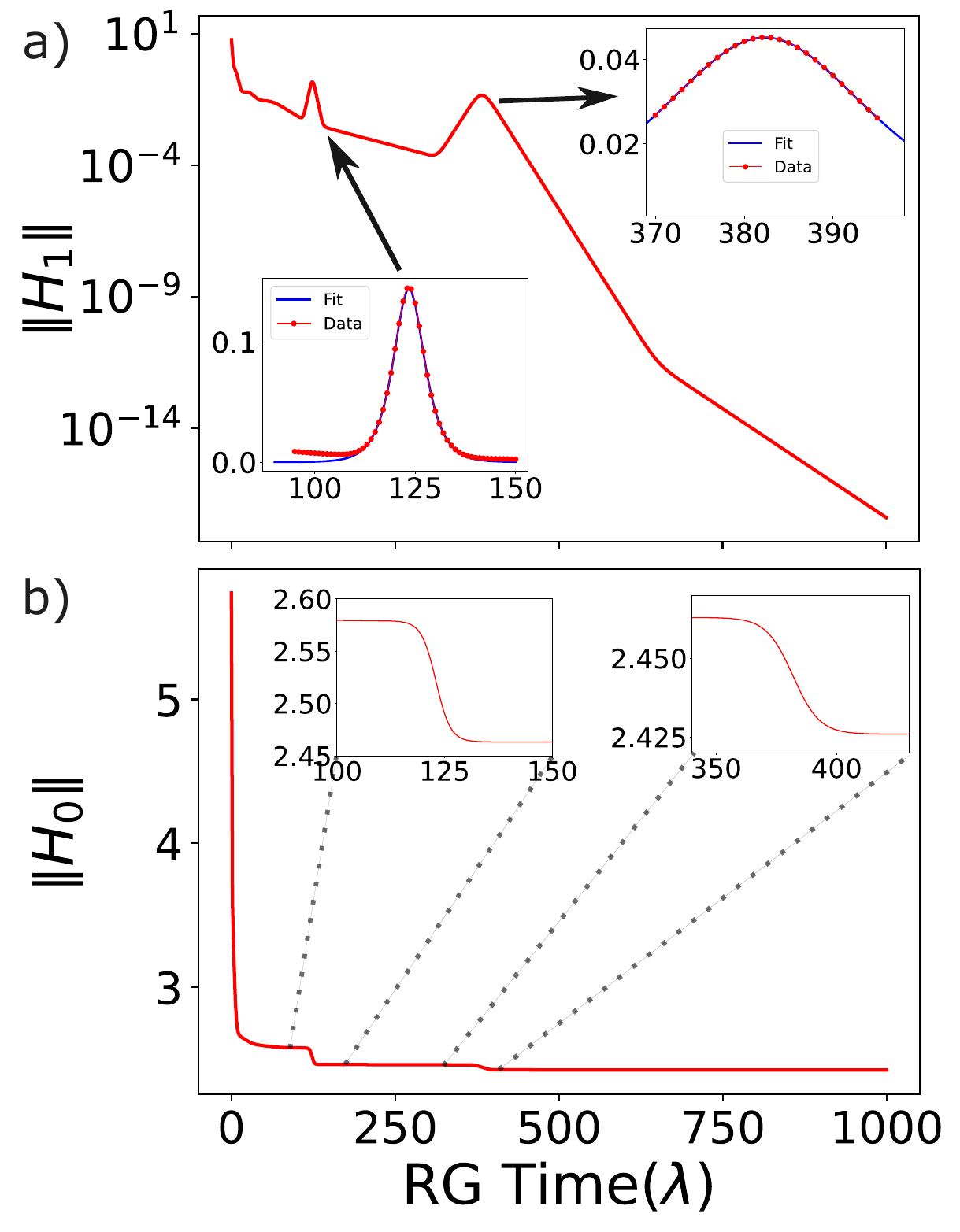}
    \caption{(a) Frobenius norm of $H_{1}(\lambda)$ as a function of fRG time at FP$_1$ for $L=6$, $\Omega=1.0J$, $J=1$,$J_{2}=0.9J$. (b) Corresponding Frobenius norm of $H_{0}(\lambda)$. The two instanton events are marked by arrows in (a), along with the corresponding step-like features in (b).}\label{fig:L6rf}
\end{figure}

\end{itemize}

\section{Early-time analysis}
\label{app:magnus}

In this appendix, we solve the flow equation in the early time and large $\Omega$ limit, and compare the result with the (modified) Magnus expansion. The Hamiltonian we consider is
\begin{equation}\label{eq:Hbare_app}
  H(t)=H_0+2H_1\cos(\Omega t)\,,
\end{equation} 
where we have chosen $H_{-1}=H_1$ and $H_1=H_1^\dagger$ in Eq.~\eqref{eq:Hbare} at $\lambda=0$.

\subsection{Review of Floquet-Magnus expansion}
To set the stage for a comparison with our flow analysis, we first perform a (modified) Floquet-Magnus expansion for Eq.~\eqref{eq:Hbare_app}, by transforming to the co-moving frame Hamiltonian. This is achieved by the unitary,
\begin{equation}\label{}
  W(t)=\exp\left(-\frac{2i H_1}{\Omega} \sin(\Omega t)\right)\,,
\end{equation} 
and the co-moving Hamiltonian is
\begin{equation}\label{}
\begin{split}
  H_\text{mov}(t) &=W(t)^\dagger \left(H(t)-i\partial_t\right) W(t) \\
  &= \exp\left(i\frac{2 \sin(\Omega t)\AdH}{\Omega}\right)H_0\,.
\end{split}
\end{equation} Here $\Ad_X Y$ is an adjoint operator acting on $Y$, i.e. $\mathrm{Ad}_X Y=[X,Y]$.

The Floquet-Magnus expansion connects $H_\text{mov}(t)$ to the effective Floquet Hamiltonian $H_F$. The first two orders of the expansion are given by
\begin{subequations}
\beq\label{}
  H_F^{(0)}&=&\frac{1}{T}\int_{0}^{T}\rd t H_\text{mov}(t)\,,\\
  H_F^{(1)}&=&\frac{1}{2iT} \int_0^T \rd t_1 \int_0^{t_1}\rd t_2[H_\text{mov}(t_1),H_\text{mov}(t_2)].\nn\\
\eeq
\end{subequations}
To evaluate the integrals, we decompose $H_\text{mov}$ into fourier components,
\begin{subequations}
\beq\label{}
  H_\text{mov}(t) &=& \sum_m h_m e^{im \Omega t}\,,~\rm{where}\\
  h_m &=&J_m\left(\frac{2 \AdH}{\Omega}\right)H_0\,,
\eeq
\end{subequations}
and $J_m$ is the Bessel function of the first kind. We thus obtain,
\begin{subequations}
\beq\label{}
  H_F^{(0)}&=&h_0\,,\\
  H_F^{(1)}&=&\sum_{m,n}I_{mn}^{(1)}[h_m,h_n]\,,
\eeq 
\end{subequations}
where the coefficient $I_{mn}^{(1)}$ is given by,
\begin{equation}\label{}
   I_{mn}^{(1)}=\frac{1}{2iT} \int_0^T \rd t_1 \int_0^{t_1}\rd t_2 \exp(i\omega(m t_1+n t_2))\,.
\end{equation}
The only nonzero values for $I_{mn}^{(1)}$ are
\begin{equation}\label{}
  I_{0m}^{(1)}=-I_{m0}^{(1)}=I_{m,-m}^{(1)}=\frac{1}{2m \Omega}\,,\quad m\neq 0\,,
\end{equation} and $I_{00}^{(1)}=-(i\pi)/(2\Omega)$. Since $[h_m,h_{-m}]=0$, we only need $I_{m0}^{(1)}$ and $I_{0m}^{(1)}$. Therefore we obtain
\begin{equation}\label{eq:Magnus1}
  H_{F}^{(1)}=2\sum_{m\neq 0} \frac{-1}{2m\Omega}[h_m,h_0]\,.
\end{equation}

\subsection{Solution of flow equation}

In this subsection, we solve the flow equation in the early-time limit, which we reproduce below, 
\begin{subequations}
\begin{eqnarray}
    \partial_\lambda H_0(\lambda)&=&2[H_1(\lambda),H_1^\dagger(\lambda)]\,, \label{eq:FlowH0_app}\\
     \partial_\lambda H_1(\lambda)&=&-\Omega H_1(\lambda)-\left[H_0(\lambda),H_1(\lambda)\right]\,,\label{eq:FlowH1_app}
\end{eqnarray} 
\end{subequations}
where the initial condition is chosen to satisfy $H_1(0)=H_1(0)^\dagger$.

We first integrate Eq.~\eqref{eq:FlowH1_app} to obtain
\begin{equation}\label{eq:H1lambda_app}
  H_1(\lambda)=e^{-\Omega \lambda} U(\lambda) H_1(0) U(\lambda)^{-1}\,,
\end{equation} where
\begin{equation}\label{}
  U(\lambda)=\mathcal{T}\exp\left(-\int_0^{\lambda} \rd \bar{\lambda}H_0(\bar{\lambda})\right),
\end{equation} where $\mathcal{T}$ denotes time-ordering in $\lambda$. Therefore, Eq.\eqref{eq:FlowH0_app} can be cast as,
\begin{equation}\label{eq:FlowH0_alt}
  \partial_\lambda H_0(\lambda)=2e^{-2\Omega \lambda}[\bar{H}_1(\lambda),\bar{H}_1(\lambda)^\dagger]\,,
\end{equation} where
\begin{subequations}
\begin{eqnarray}
  \bar{H}_1(\lambda) &=& U(\lambda) H_1(0)U(\lambda)^{-1}, \label{eq:barH1} \\
  \bar{H}_1(\lambda)^\dagger &=& U(\lambda)^{-1} H_1(0)^\dagger U(\lambda)\,. \label{eq:barH1d}
\end{eqnarray} 
\end{subequations}
Taking an additional derivative in Eq.\eqref{eq:FlowH0_alt},  we obtain a (nonlinear) second-order differential equation for $H_0(\lambda)$:
\beq\label{}\label{eq:H0_diff}
  &&\partial_\lambda^2 H_0(\lambda)+2\Omega \partial_\lambda H_0(\lambda)+\nn\\
 && 2 e^{-2\Omega \lambda} \left\{\Ad_{\bar{H}_1(\lambda)},\Ad_{\bar{H}_{1}(\lambda)^\dagger}\right\}H_0(\lambda)=0\,.
\eeq Here $\bar{H}_1(\lambda)$ and $\bar{H}_1(\lambda)^\dagger$ depend on $H_0(\lambda)$, as described by Eqs.~\eqref{eq:barH1}, and \eqref{eq:barH1d}, and $\{\cdot,\cdot\}$ denotes the usual anticommutator.

We emphasize that all of the manipulations thus far are exact. To make connection with the Floquet-Magnus expansion above, we can expand  Eq.~\eqref{eq:H0_diff} in powers of $H_0(0)$, i.e.
\begin{equation}\label{}
  H_0(\lambda)=\zeta H_0^{(1)}(\lambda)+\zeta^2 H_0^{(2)}(\lambda)+\zeta^3 H_0^{(3)}(\lambda)+\dots\,,
\end{equation} where $H_0^{(k)}(\lambda)$ depends on the $k$-th power of $H_0(0)$, and $\zeta$ is a convenient book keeping parameter to organize the expansion. We now obtain the forms of these corrections up to the third order term.

\begin{itemize}[leftmargin=*]
    \item {\bf First order term:} We first solve for $H_0^{(1)}$, which is obtained by setting $\bar{H}_1(\lambda)=\bar{H}_1(\lambda)^\dagger=H_1(0)$ in Eq.~\eqref{eq:H0_diff}, and the associated linear second-order differential equation:
\begin{equation}\label{}
  \partial_\lambda^2 H_0^{(1)}(\lambda)+2\Omega \partial_{\lambda} H_0^{(1)}(\lambda)+4 e^{-2\Omega \lambda}\mathrm{Ad}_{H_1(0)}^2 H_0^{(1)}(\lambda)=0\,.
\end{equation}
The analytical solution, with an initial condition
 $H_0^{(1)}(0)=H_0(0)$, $\partial_\lambda H_0^{(1)}(0)=0$, is given by
\begin{equation}\label{eq:H0lambda_app}
  H_0^{(1)}(\lambda)=\frac{\pi}{2}e^{- \Omega\lambda} \hat{z}\left[J_1(\hat{z}e^{-\lambda \Omega})Y_0(\hat{z})-Y_1(\hat{z}e^{-\Omega\lambda})J_0(\hat{z})\right]H_0(0)\,,
\end{equation} where $\hat{z}=2\mathrm{Ad}_{H_1(0)}/\Omega$, and $Y_m$ denotes Bessel function of the second kind. Taking the $\lambda\to\infty$ limit, we obtain a result consistent with the leading order magnus expansion:
\begin{equation}\label{}
  H_0^{(1)}(\lambda\to\infty)=J_0\left(\frac{2\mathrm{Ad}_{H_1(0)}}{\Omega}\right)H_0(0)\,.
\end{equation}

We apply the results above to the driven Harmonic oscillator example discussed in Sec.~\ref{sec:HO}.  It implies that the $\rm{U(1)}$ violating component $B_0(\lambda)$ of $H_0(\lambda)$, in the $\omega_0\ll \Omega$ limit, will take the form 
    \beq\label{}
    B_0(\lambda)&=&\frac{\pi A \omega_1}{2\Omega}e^{-\Omega\lambda}\bigg[J_1\bigg(\frac{A}{\Omega}e^{-\Omega\lambda}\bigg)Y_0\bigg(\frac{A}{\Omega}\bigg)\nn\\
    &-&J_0\bigg(\frac{A}{\Omega}\bigg) Y_1\bigg(\frac{A}{\Omega}e^{-\Omega \lambda}\bigg)\bigg].
  \eeq
Then the $\lambda\to\infty$ limit reads $B_0(\lambda\to\infty)=\omega_1 J_0(A/\Omega)$. Therefore, to leading order the $\rm{U(1)}$ symmetry emerges when $A/\Omega$ equals zeros of the Bessel function $J_0$, which agrees with prediction of the Magnus expansion.

\item {\bf Second order term:} The term, $H_0^{(2)}(\lambda)$, can be obtained by expanding Eq.~\eqref{eq:H0_diff} to ${\cal{O}}(\zeta^2)$, with a resulting equation:
\begin{equation}\label{}
  \partial_\lambda^2 H_0^{(2)}(\lambda)+2\Omega \partial_{\lambda} H_0^{(2)}(\lambda)+4 e^{-2\Omega \lambda}\mathrm{Ad}_{H_1(0)}^2 H_0^{(2)}(\lambda)=S^{(2)}(\lambda)\,.
\end{equation} 
The source term, $S^{(2)}(\lambda)$, is obtained by expanding the last term of Eq.~\eqref{eq:H0_diff} to ${\cal{O}}(\zeta^2)$,
\beq\label{eq:S2}\label{eq:source}
  S^{(2)}(\lambda)=-2e^{-2\Omega\lambda}\bigg[\bigg\{\Ad_{\bar{H}_1^{(1)}(\lambda)},\Ad_{\bar{H}_1^{(0)}(\lambda)^\dagger}\bigg\} \nn\\
  +\bigg\{\Ad_{\bar{H}_1^{(0)}(\lambda)},\Ad_{\bar{H}_1^{(1)}(\lambda)^\dagger}\bigg\}\bigg]H_0^{(1)}(\lambda)\,.
\eeq
Here we have similarly expanded $\bar{H}_1(\lambda)$ in powers of $\zeta$:
\begin{equation}\label{}
  \bar{H}_1(\lambda)=\bar{H}_1^{(0)}(\lambda)+\zeta \bar{H}_1^{(1)}(\lambda)+\dots\,,
\end{equation} and the leading order term is just the bare drive Hamiltonian $\bar{H}_1^{(0)}(\lambda)=H_1(0)$. The first order term is obtained by expanding Eq.~\eqref{eq:barH1},
\begin{equation}\label{eq:barH1_1}
  \bar{H}_1^{(1)}(\lambda)=-\int_0^\lambda\rd\bar{\lambda}\left[H_0^{(1)}(\bar{\lambda}),H_1(0)\right]\,.
\end{equation} 
Interestingly, due to conjugation, $\bar{H}_1^{(1)}(\lambda)^\dagger = -\bar{H}_1^{(1)}(\lambda)$, the two terms in Eq.~\eqref{eq:S2} cancel, and hence the source term, $S^{(2)}(\lambda)=0$, in Eq.~\ref{eq:source} vanishes. Combining with the initial condition $H_0^{(2)}(0)=\partial_\lambda H_0^{(2)}(0)=0$, the second order term vanishes identically.

\item {\bf Third order term:} The first non-trivial correction appears at the third order, which satisfies
\begin{equation}\label{eq:H03_diff}
  \partial_\lambda^2 H_0^{(3)}(\lambda)+2\Omega \partial_{\lambda} H_0^{(3)}(\lambda)+4 e^{-2\Omega \lambda}\mathrm{Ad}_{H_1(0)}^2 H_0^{(3)}(\lambda)=S^{(3)}(\lambda)\,,
\end{equation} 
with a source term,
\beq\label{}
  S^{(3)}(\lambda)&=&-2e^{-2\Omega\lambda}\bigg[\bigg\{\Ad_{\bar{H}_1^{(2)}(\lambda)},\Ad_{\bar{H}_1^{(0)}(\lambda)^\dagger}\bigg\} \nn\\
  &+&\bigg\{\Ad_{\bar{H}_1^{(0)}(\lambda)},\Ad_{\bar{H}_1^{(2)}(\lambda)^\dagger}\bigg\} \nn\\
  &+&2\bigg\{\Ad_{\bar{H}_1^{(1)}(\lambda)},\Ad_{\bar{H}_1^{(1)}(\lambda)^\dagger}\bigg\}\bigg]H_0^{(1)}(\lambda)\,.
\eeq
Here, the second order term  $\bar{H}_1^{(2)}(\lambda)$ is
\begin{equation}\label{eq:barH1_2}
  \bar{H}_1^{(2)}(\lambda)=\int_{0}^{\lambda} \rd \bar{\lambda}_1 \int_0^{\bar{\lambda}_1}\rd\bar{\lambda}_2 \left[H_0^{(1)}(\bar{\lambda_1}),\left[H_0^{(1)}(\bar{\lambda_2}),H_1(0)\right]\right]\,.
\end{equation}
The source term, $S^{(3)}$, is in general nonzero. Combining Eq.~\eqref{eq:H03_diff} with the initial condition $H_0^{(3)}(0)=\partial_\lambda H_0^{(3)}(0)=0$, we can in principle calculate the correction to $H_0(\lambda\to\infty)$. Instead of carrying out this computation explicitly, we perform dimensional analysis of the result. Since the expansion here is in powers of $H_0$, an additional energy scale must be included, which turns out to be the driving frequency $\Omega$. This can be seen by noting that each integral in $\lambda$ (Eqs.~\eqref{eq:barH1_1},~\eqref{eq:barH1_2}) introduces a factor of $1/\Omega$, due to the fact that $\lambda$ appears in Eq.~\eqref{eq:H0lambda} in the form of $\Omega\lambda$. Therefore, we expect $H_0^{(3)}(\lambda)$ to take the form:
\begin{equation}\label{}
  H_0^{(3)}(\lambda)\sim \frac{1}{\Omega^2}\hat{\mathcal{F}}\left(\frac{H_1(0)}{\Omega}\right)\left[H_0(0)\otimes H_0(0)\otimes H_0(0)\right],
\end{equation} where $\hat{\mathcal{F}}$ is a linear functional that acts on $H_0(0)\otimes H_0(0)\otimes H_0(0)$. This is the promised $1/\Omega^2$ correction, which we have found numerical evidence for previously in Fig.~\ref{fig:EEmag}. Furthermore, we find that the flow formalism agrees with the modified Magnus expansion at the leading order, but the subsequent correction terms do not agree, as described previously using arguments relying on gauge-invariance.

\end{itemize}

\section{Estimate of thermalization timescales}\label{app:timescale}

In this section, we provide an estimate for the thermalization time scale of the Floquet system in the limit of a large drive frequency, $\Omega$. Our discussion will mainly involve a spin chain as an illustrative example, where the typical energy cost of a single spin flip is denoted $J$. We also assume that the amplitude $A$ of the drive $H_1(0)$ is at most of the same order as the frequency $\Omega$, and both $H_0(0)$ and $H_1(0)$ are sum of local terms.  As discussed in the main text, thermalization happens in the form of instantons, which corresponds to a matrix element of $H_1(\lambda)$ that flips $k$ spins at the same time, with $k>\mathcal{O}{(\Omega/J)}$. Our approach relies on first flowing to the prethermal regime, and estimating the typical matrix element, $H_1(\lambda_\text{pre})$, that flips $k$ spins. Next, we estimate the fRG time $\lambda_\text{min}$ where $\Fnorm{H_1(\lambda)}$ is minimized. Finally, we translate this to a real time scale using the relation $t_\text{eff}\sim\Fnorm{H_1(\lambda_\text{min})}^{-1}$.

Let us estimate the typical size of the matrix element, $\braket{m|H_1(\lambda_\text{pre})|n}$, which flips $k$ spins. From Eq.~\eqref{eq:H0lambda_app}, we see that $H_0(\lambda)$ will converge to its prethermal limit in fRG time of order a few drive cycles $\lambda\sim \Omega^{-1}$. Therefore, we can choose a $\lambda_\text{pre}$ so that $\Omega \lambda_{\text{pre}}=\zeta$. Here we choose $\zeta\gg 1$ to ensure we are close to the prethermal fixed point, but $\zeta$ is still parametrically smaller than the thermalization time scale, i.e. $\zeta$ is an $\mathcal{O}(1)$ number  that does not diverge in the $\Omega/J\to\infty$ limit with $A/\Omega$ kept fixed. This assumption is verified {\it a posteriori} by the fact that $H_0(\lambda)$ approaches the zeroth order magnus result faster at larger $\Omega$ (see Fig.~\ref{fig:EEmag} (b)). Assuming now $H_0\approx H_0(\lambda_\text{pre})$ , we can approximate $U(\lambda)\approx \exp(-\lambda H_0(\lambda_\text{pre}))$ in Eq.~\eqref{eq:H1lambda_app}. Expanding $H_1(\lambda_\text{pre})$ in Eq.~\eqref{eq:H1lambda_app} using the BCH formula, 
\begin{equation}\label{}
  H_1(\lambda_\text{pre})=e^{-\Omega\lambda_\text{pre}}\sum_k \frac{(-1)^k\lambda_\text{pre}^k}{k!}\Ad_{H_0(\lambda_\text{pre})}^k H_1(0)\,. 
\end{equation} 
The commutator with $H_0(\lambda_\text{pre})$ will increase the range of operators by $\mathcal{O}(1)$. Therefore, terms involving $k$ spin flips are from order $\mathcal{O}(k)$ of the expansion. We replace $\lambda_\text{pre}^k$ by $\zeta^k \Omega^{-k}$. Using $\Fnorm{\Ad_{H_0(\lambda_\text{pre})}^k H_1(0)}\leq J^k k!$ \cite{prethermal1}, the matrix element that involves $k$ spin flips will be bounded by order $e^{-\zeta}(\zeta J/\Omega)^k$. In the following analysis, we are interested in the asymptotic limit of $\Omega/J\gg1$, and we therefore drop the $\zeta$ factor which does not grow with $\Omega/J$.

Now, let us resume the flow and analyze the evolution of $\Fnorm{H_1(\lambda)}$. As discussed in the main text, matrix element of $H_1(\lambda)$ with $k$ spin flips will evolve with the factor $\exp[(kJ-\Omega)(\lambda-\lambda_\text{pre})]$. Summing contributions of different $k$ to $\Fnorm{H_1(\lambda)}$, we obtain
\begin{equation}\label{eq:H1termbyterm}
  \Fnorm{H_1(\lambda)}\sim \sum_k a_k \left(\frac{ J}{\Omega}\right)^k e^{(\lambda-\lambda_\text{pre})(kJ-\Omega)}\,.
\end{equation} Here $a_k$ are numerical coefficients that we expect to decay to zero as $k\to\infty$. 
 
\begin{figure}[t]
    \centering
    \includegraphics[width=1.0\linewidth]{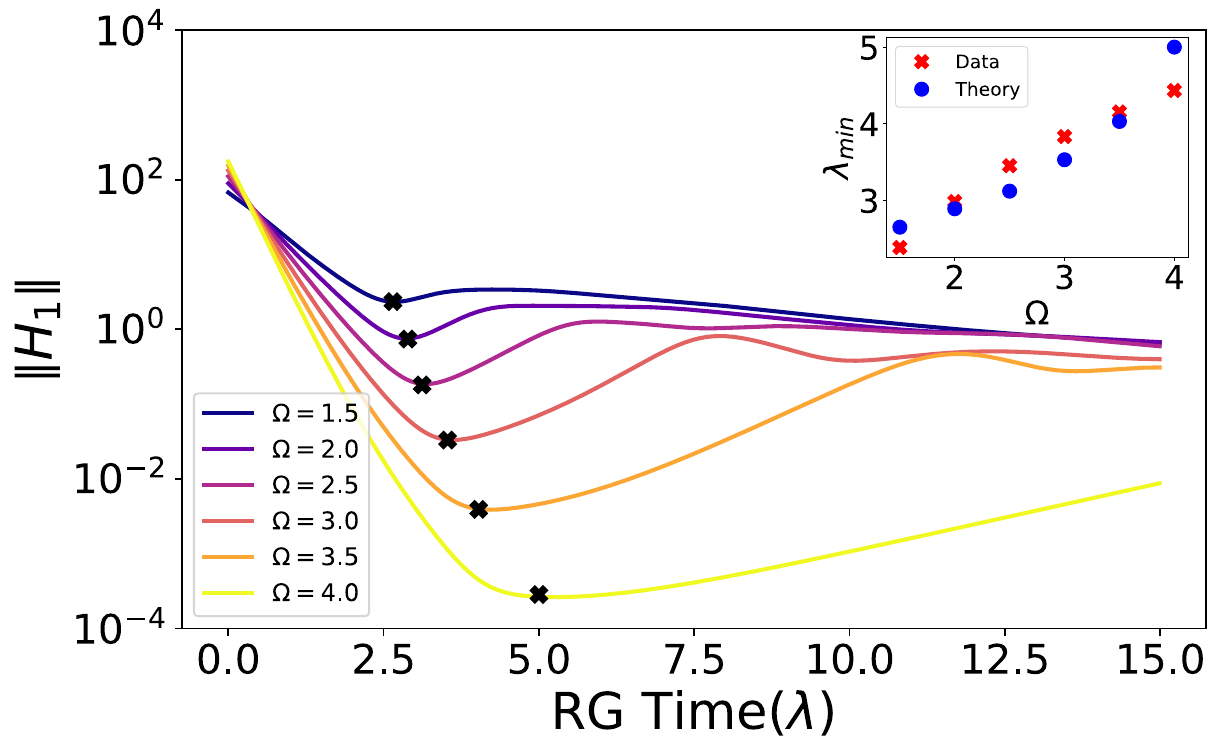}
    \caption{Driving frequency ($\Omega$) scaling of $H_{1}(\lambda)$ as a function of fRG time for $J=1$, $J_{2}=0.9$, $L=10$, and $A/\Omega=0.9$. Inset: Value of $\lambda$ corresponding to the minimum norm of $H_{1}$ as a function of $\Omega$, demonstrating an order-of-magnitude agreement with the theoretical prediction in Eq.~\eqref{Eq:lambdamin}. Here, the theoretical data points are obtained by substituting $J=0.47$ in Eq.~\eqref{Eq:lambdamin}, which is roughly the spin-flip energy associated with the non-integrable coupling $J_2$.}\label{fig:L10thermalization}
\end{figure}

Since $\Omega\gg J$, $\Fnorm{H_1(\lambda)}$ will initially show decay for small $(\lambda-\lambda_\text{pre})$, followed by a growth for larger $(\lambda-\lambda_\text{pre})$. To estimate the minimum time $\lambda_\text{min}$, we rewrite $k=\Omega/J+\delta k$, and we have 
\begin{equation}\label{}
  \Fnorm{H_1(\lambda)}\sim \sum_{\delta k} a_k \left(\frac{J}{\Omega}\right)^{\delta k +\Omega/J} e^{(\lambda-\lambda_\text{pre}) \delta k J}\,.
\end{equation} Since $\Omega/J\gg 1$, the dependence of $a_k$ on $\delta k$ will be weak, and we approximate the summation by $\delta k$ from $-\infty$ to $\infty$. The minimum is reached when the terms of exponent $\delta k$ are balanced by terms of exponent $-\delta k$, meaning that $(J/\Omega)\exp(J(\lambda-\lambda_\text{pre}))\sim\mathcal{O}(1)$. Therefore, we obtain 
\begin{equation}\label{Eq:lambdamin}
  \lambda_\text{min}-\lambda_\text{pre}\sim \frac{1}{J}\ln\frac{\Omega}{J}\,.
\end{equation} The dependence of $a_k$ on $k$ does not alter the estimate significantly. For example, if $a_k\propto (k!)^{-\alpha}$, it only changes the numerical prefactor of the estimate due to $k\sim \Omega/J$.  At the same time, $\Fnorm{H_1}$ is roughly,
\begin{equation}\label{}
  \Fnorm{H_1(\lambda_\text{min})}\sim \left(\frac{J}{\Omega}\right)^{\Omega/J} \sim \exp\left(-\frac{\Omega}{J}\ln\frac{\Omega}{J}\right)\,.
\end{equation} Therefore, the heating rate is exponentially small in $\Omega$, and our estimate is consistent with \cite{prethermal1}. We also see that our choice of $\Omega\lambda_\text{pre}\gg 1$ is consistent as $\lambda_\text{min}-\lambda_\text{pre}\gg 1/\Omega$. In Fig.~\ref{fig:L10thermalization} we plot the $\Omega$ scaling of the norm of $H_{1}(\lambda)$ which shows order of magnitude agreement with our theoretical prediction.

Another fRG timescale $\lambda_\text{th}$ describes the time when $\Fnorm{H_1(\lambda)}$ grows to order one. Each term of Eq.~\eqref{eq:H1termbyterm} will yield a different estimate. For $k=l$, we obtain 
\begin{equation}
     \lambda^{(l)}_\text{th}\sim \frac{f(l)+l\ln(\Omega/J)}{lJ-\Omega}\,.
\end{equation} Here $f(l)=\ln(1/a_l)$, and is expected to be a smooth function. For example, $a_l\propto (l!)^{-\alpha}$ will lead to $f(l)\sim \alpha l\ln l$. The estimate above shows that only operators that flip $l>\mathcal{O}(\Omega/J)$ spins will lead to growth of $H_1(\lambda)$, and hence an instanton.  The minimum of $\lambda_\text{th}^{(l)}$ is expected to be parametrically similar to $\lambda_\text{min}$. 

\section{Floquet-Magnus analysis for driven spin chain}\label{app:Magnus_spinchain}

        We compute the Floquet-Magnus expansion for the spin-chain Hamiltonian, utilizing results derived in Appendix~\ref{app:magnus}. The unitary transformation to the rotating frame is
        \begin{equation}\label{}
          W(t)=\exp\left(-i\frac{\alpha}{2} \sin(\Omega t) \sum_i S_i^x\right)\,,
        \end{equation} where $\alpha=4A/\Omega$.

        The co-moving Hamiltonian is
\begin{widetext}
\begin{equation}\label{}
        H_\text{mov}(t)=-\sum_{d=1,2}\frac{J^{(d)}}{2}\sum_i\left\{S^y_i S^y_{i+d}\left[1-\cos\left(\alpha\sin(\Omega t)\right)\right]+S^z_i S^z_{i+d}\left[1+\cos\left(\alpha\sin(\Omega t)\right)\right]\right.
          +\left. \left[S^y_i S^z_{i+d}+S^z_i S^y_{i+d}\right]\sin(\alpha\sin(\Omega t))\right\}\,,
\end{equation}
where $J^{(1)}=J$ and $J^{(2)}=J_2$. Expanding into fourier components, $h_m=\frac{1}{T}\int_0^T H_\text{mov}(t)e^{-im\Omega t}$, we obtain
    \begin{align}\label{Eq:effective}
        h_m=-\sum_{d=1,2}\frac{J^{(d)}}{2}\sum_i\left\{S^y_i S^y_{i+d}\left[\delta_{m,0}-\frac{1+(-1)^m}{2}J_m(\alpha)\right]+S^z_i S^z_{i+d}\left[\delta_{m,0}+\frac{1+(-1)^m}{2}J_m(\alpha)\right]\right.\\
        +\left. \left[S^y_i S^z_{i+d}+S^z_i S^y_{i+d}\right]\frac{1-(-1)^m}{2i}J_m(\alpha)\right\}\,,
    \end{align}
\end{widetext}
The zeroth Floquet Hamiltonian is therefore
        \begin{equation}\label{}
          H_F^{(0)}=h_0\,.
        \end{equation} When $J_0(\alpha)=0$, $H_F^{(0)}\sim \sum_i (S_i^y S_{i+1}^y+S_i^z S_{i+1}^z)+ \sum_i (S_i^y S_{i+2}^y+S_i^z S_{i+2}^z)$, which commutes with $\sum_i S_i^x$. The result above is compared with the flow Hamiltonian $H_0(\lambda)$ in Fig.~\ref{fig:EEmag}, where the difference between the two decays as $1/\Omega^2$ for large $\Omega$, implying that the leading order result is the same.

\bibliography{Floquet}

\newpage

\clearpage

\foreach \x in {1,...,4}
{
\clearpage
\includepdf[pages={\x},angle=0]{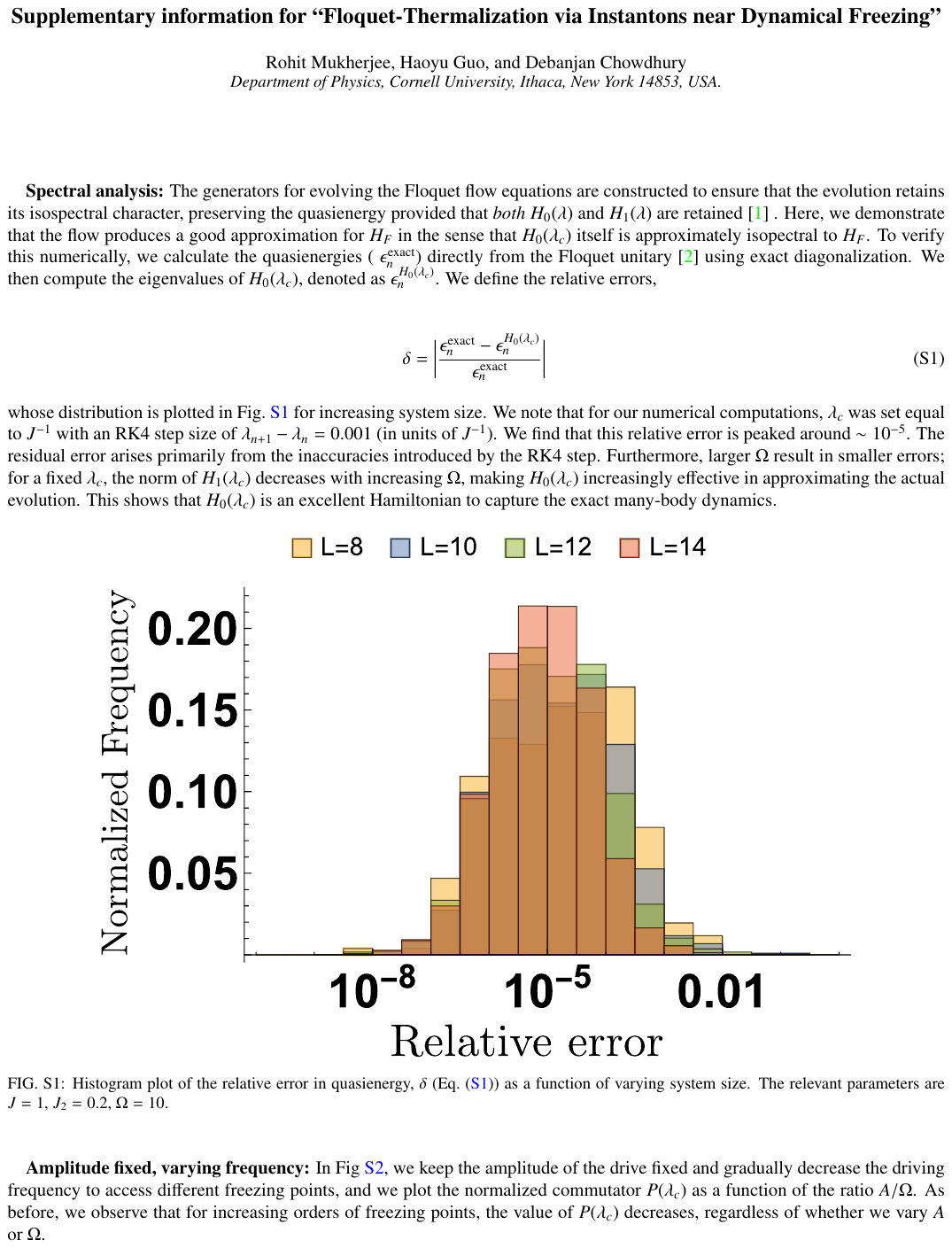}
}

\end{document}